\documentclass[aps,twocolumn,prx,floatfix,lengthcheck,citeautoscript,superscriptaddress,longbibliography]{revtex4-2}
\usepackage{amssymb,amsfonts,amsmath,amsthm}
\usepackage{graphicx}%esto va sin el DVIPS!
\usepackage[english]{babel}
\usepackage[latin1]{inputenc}
\usepackage[pdftex,colorlinks=true,pdfstartview=FitH,linkcolor=blue,citecolor=blue,urlcolor=blue]{hyperref}

\usepackage{bm}
\usepackage{float}
\usepackage{mathtools}
\usepackage{color,soul}
\usepackage{times}
\usepackage{upgreek}
\usepackage{MnSymbol}
\usepackage{verbatim}
\usepackage[bottom]{footmisc}
\usepackage{bbold}

\usepackage[dvipsnames,svgnames,table]{xcolor}
\usepackage{hyperref}
\usepackage[english]{babel}
\usepackage[caption=false]{subfig}

\hypersetup{
pdfstartview={FitH},% fits the width of the page to the window
colorlinks=true,    % false: boxed links; true: colored links
linkcolor=NavyBlue, % color of internal links
citecolor=Blue,   % color of links to bibliography
filecolor=NavyBlue, % color of file links
urlcolor=NavyBlue   % color of external links
}

\newcommand{\bea}{\begin{eqnarray}}
\newcommand{\eea}{\end{eqnarray}}
\newcommand{\be}{\begin{equation}}
\newcommand{\ee}{\end{equation}}

\newcommand{\ci}{\mathrm{i}}

\begin{document}
%%%%%%%%%%%%%%%%%%%%%%%%%%%%%%%%%%%%%%%%%%%%%%%%%%%%%%%%%%%%%%
\title{
%A quantum cascade phonon laser \\ 
%or \\
%Bosonic cascade phonon laser \\ 
%or  \\
Polariton cascade phonon laser
}
%%%%%%%%%%%%%%%%%%%%%%%%%%%%%%%%%%%%%%%%%%%%%%%%%%%%%%%%%%%%%

\author{I. Papuccio-Fern\'andez}
\affiliation{Centro At{\'{o}}mico Bariloche and Instituto Balseiro,
Comisi\'on Nacional de Energ\'{\i}a At\'omica (CNEA)- Universidad Nacional de Cuyo (UNCUYO), 8400 Bariloche, Argentina.}
\affiliation{Instituto de Nanociencia y Nanotecnolog\'{i}a (INN-Bariloche), Consejo Nacional de Investigaciones Cient\'{\i}ficas y T\'ecnicas (CONICET), Argentina.}

%\author{I. Carraro Haddad}
%\affiliation{Centro At{\'{o}}mico Bariloche and Instituto Balseiro, Comisi\'on Nacional de Energ\'{\i}a At\'omica (CNEA)- Universidad Nacional de Cuyo (UNCUYO), 8400 Bariloche, Argentina.}
%\affiliation{Instituto de Nanociencia y Nanotecnolog\'{i}a (INN-Bariloche), Consejo Nacional de Investigaciones Cient\'{\i}ficas y T\'ecnicas (CONICET), Argentina.}

%\author{D. L. Chafatinos}
%\thanks{These two authors contributed equally}
%\affiliation{Centro At{\'{o}}mico Bariloche and Instituto Balseiro, Comisi\'on Nacional de Energ\'{\i}a At\'omica (CNEA)- Universidad Nacional de Cuyo (UNCUYO), 8400 Bariloche, Argentina.}
%\affiliation{Instituto de Nanociencia y Nanotecnolog\'{i}a (INN-Bariloche), Consejo Nacional de Investigaciones Cient\'{\i}ficas y T\'ecnicas (CONICET), Argentina.}

\author{A.~A. Reynoso}
\affiliation{Centro At{\'{o}}mico Bariloche and Instituto Balseiro,
Comisi\'on Nacional de Energ\'{\i}a At\'omica (CNEA)- Universidad Nacional de Cuyo (UNCUYO), 8400 Bariloche, Argentina.}
\affiliation{Instituto de Nanociencia y Nanotecnolog\'{i}a (INN-Bariloche), Consejo Nacional de Investigaciones Cient\'{\i}ficas y T\'ecnicas (CONICET), Argentina.}
%\affiliation{Departamento de F\'isica Aplicada II, Universidad de Sevilla, E-41012 Sevilla, Spain}

\author{A. E. Bruchhausen}
\affiliation{Centro At{\'{o}}mico Bariloche and Instituto Balseiro,
Comisi\'on Nacional de Energ\'{\i}a At\'omica (CNEA)- Universidad Nacional de Cuyo (UNCUYO), 8400 Bariloche, Argentina.}
\affiliation{Instituto de Nanociencia y Nanotecnolog\'{i}a (INN-Bariloche), Consejo Nacional de Investigaciones Cient\'{\i}ficas y T\'ecnicas (CONICET), Argentina.}

\author{A.~S. Kuznetsov}
\affiliation{Paul-Drude-Institut f\"{u}r Festk\"{o}rperelektronik, Leibniz-Institut im Forschungsverbund Berlin e.V., Hausvogteiplatz 5-7,\\ 10117 Berlin, Germany.}

\author{K. Biermann}
\affiliation{Paul-Drude-Institut f\"{u}r Festk\"{o}rperelektronik, Leibniz-Institut im Forschungsverbund Berlin e.V., Hausvogteiplatz 5-7,\\ 10117 Berlin, Germany.}

\author{P.~V. Santos}
\email[Corresponding author, e-mail: ]{santos@pdi-berlin.de}
\affiliation{Paul-Drude-Institut f\"{u}r Festk\"{o}rperelektronik, Leibniz-Institut im Forschungsverbund Berlin e.V., Hausvogteiplatz 5-7,\\ 10117 Berlin, Germany.}

\author{G. Usaj}
%\email[Corresponding author, e-mail: ]{gonzalo.usaj@ib.edu.ar}
\affiliation{Centro At{\'{o}}mico Bariloche and Instituto Balseiro,
Comisi\'on Nacional de Energ\'{\i}a At\'omica (CNEA)- Universidad Nacional de Cuyo (UNCUYO), 8400 Bariloche, Argentina.}
\affiliation{Instituto de Nanociencia y Nanotecnolog\'{i}a (INN-Bariloche), Consejo Nacional de Investigaciones Cient\'{\i}ficas y T\'ecnicas (CONICET), Argentina.}

\author{A. Fainstein}
\email[Corresponding author, e-mail: ]{afains@cab.cnea.gov.ar}
\affiliation{Centro At{\'{o}}mico Bariloche and Instituto Balseiro,
Comisi\'on Nacional de Energ\'{\i}a At\'omica (CNEA)- Universidad Nacional de Cuyo (UNCUYO), 8400 Bariloche, Argentina.}
\affiliation{Instituto de Nanociencia y Nanotecnolog\'{i}a (INN-Bariloche), Consejo Nacional de Investigaciones Cient\'{\i}ficas y T\'ecnicas (CONICET), Argentina.}

\date{\today}

\begin{abstract}
Phonon lasers, as their photon counterparts, rely on the physics of stimulated emission. Arguably, because light does not require a material substrate to propagate, while sound does, the impact of the two technologies has however been highly contrasting, with ``sasers'' (for sound amplification by stimulated emission of radiation) mostly remaining as an academic curiosity. This might be changing due to the possibility to use coherent sound generation for on-chip processing of information at ultra-high frequencies, and in the quantum realm, in integrated photonic and optomechanical devices. Inspired by the concept of unipolar lasers based on the quantum engineering of states in semiconductor heterostructures, we propose and implement a quantum cascade phonon laser (QCPL). A condensate of exciton-photon  quasiparticles (polaritons) is optically induced in a microstructured semiconductor device to jump down a ladder of engineered levels. This down-cascade is accompanied by the efficient stimulated emission of phonons of $\sim 20$, $\sim 60$, and $\sim 100$~GHz, which are designed to strongly interact with the polaritons on the same chip. The proposed concept opens the path for the design of integrated high-frequency optomechanical devices, as for example for non-reciprocal photon transport and multi-wavelength Brillouin lasers.
\end{abstract}
%%%%%%%%%%%%%%%%%%%%%%%%%%%%%%%%%%%%%%%%%%%%%
%\pacs{63.22.+m,78.30.Fs,78.30.-j,78.67.Pt}
\maketitle

%%%%%%%%%%%%%%%%%%%%%%%
\section{Introduction}
%%%%%%%%%%%%%%%%%%%%%%
The development of phonon lasers (or ``sasers'' for the stimulated emission of sound instead of light) was born practically at the same time as their photon counterparts. Indeed, within a year of the first report of lasing in ruby, amplification of GHz ultrasound pulses based on an inverted spin population was demonstrated on the same material~\cite{Tucker1961}. More than three decades later, an insightful series of experiments showed that the inverted spin population could be optically prepared, and the process stimulated by the phonon cavity formed by the end faces of a ruby crystal~\cite{Fokker1997a,Fokker1997b}. True phonon lasing was however not attained, and the attention migrated to solid state devices following similar developments in the field of lasers.

Solid state lasers rely on the recombination of electrons and holes across the energy gap that separates  the conduction and valence bands of a bulk semiconductor. Instead, a quantum cascade laser (QCL)~\cite{Faist1994} is a semiconductor device involving only one type of carriers. The photon emission is obtained by exploiting the so-called inter-subband transitions, where electrons tunnel between quantum confined states in traps created by specifically engineered sequences of ultra thin alternating layers of semiconductor materials. The idea to exploit such cascade of carriers along confined states in multiple quantum wells to generate phonons, instead of infrared photons as in QCLs, was  theoretically proposed many years ago in Ref.~\cite{Makler1998}, and later implemented in \cite{Beardsley2010},  including the use of a layered phonon cavity~\cite{Trigo2002} for feedback. However, the demonstrated gain was marginal, probably due to limitations in the vertical transport of charged fermions across tunnel-connected traps, and due to the limited efficiency of acoustic phonon emission when compared to that of photons in transitions involving polar carriers.

In QCLs the cascade is produced by fermions, and thus final state stimulation only relies on the emitted photons. Double stimulation can be attained, however, if the transitions also involve bosonic particles, as is the situation encountered in Brillouin lasers. The connection between stimulated Brillouin scattering, and the concomitant generation of hypersonic waves, was indeed identified from the beginning of the field~\cite{Chiao1964}. In more recent years, this approach has been developed up to exquisite levels accompanying the development of compact resonators in integrated photonics~\cite{Kharel2019,Enzian2019,Gundavarapu2019,Kuznetsov2023}. Sharing conceptual similarities with Brillouin lasers, the field of cavity optomechanics~\cite{AspelmeyerRMP2014} has also led to the demonstration of very efficient mechanical self oscillation~\cite{Carmon2005,Grudinin2010}. Central to cavity optomechanics, and distinct to the earlier field of Brillouin stimulated emission, is the concept of dynamical feedback which involves the simultaneous confinement and non-linear interaction between photons and phonons in resonators.  
Some recent ideas in this field involve non-Hermitian physics with dissipative coupling, PT-symmetries and exceptional points, to increase efficiency and to demonstrate non reciprocity~\cite{Poshakinskiy2016,Zhang2018,Jiang2018}. It is to be noted that in most of these more recent developments the stimulated vibrational modes correspond to localized oscillations, as for example in a micro-ring, a nanobeam, or a levitated nanoparticle. This is somewhat different from the concept of a laser, in which a coherent beam of bosons is emitted to be redirected and used out of the device itself.  

In this work we draw inspiration from these previous developments, and particularly from the paradigmatic concept of QCLs, to demonstrate its phonon counterpart: the quantum cascade phonon laser (QCPL). As in Brillouin lasers, we use transitions involving bosonic particles, here exciton-polaritons (shortly, polaritons)~\cite{CarusottoRMP2013}, together with a specifically engineered ladder of confined polariton energy levels, to attain a cascade of bosonic polariton condensates~\cite{Kasprzak2006,Amo2009}. We note that this approach is intimately related to the theoretical proposal of bosonic cascade lasers as a mean to generate THz electromagnetic radiation~\cite{Liew2013,Liew2016,Kaliteevski2014}, and more precisely to the double resonant approach in which both bosonic fields (polaritons and photons in the proposed THz lasers, and polaritons and phonons in our case) become stimulated~\cite{Kaliteevski2014}.

The energy levels of polaritons can be engineered based on semiconductor materials using different technologies, including deep etching, spatially modulated illumination, or cavity-spacer microstructuring~\cite{Schneider2016,Bajoni2008,Galbiati2012,Askitopoulos2013,Cristofolini2013,Alyatkin2021,Winkler2015,Kuznetsov2018}. Being neutral particles, polaritons  transition accompanied by the emission of acoustic phonons in Brillouin-like processes that can lead to stimulated phonon emission~\cite{Yulin2019,Chafatinos2020,Zambon2022}. It has been shown that the efficiency of such process can be optimized to attain the so-called strong-coupling regime (i.e., a transition rate larger than the decay rate of the involved polaritons and phonons)~\cite{Kuznetsov2023,AspelmeyerRMP2014,Sesin2023}. Moreover, the remaining Coulomb interactions, instead of leading to transport blockade as occurs for fermionic particles, can contribute positively by facilitating the locking of polariton neighbor states to the required energy of coherently emitted phonons (a concept related to synchronization)~\cite{Chafatinos2023,Ramos2024}. We show that all together, these ingredients can be combined to demonstrate an operational QCPL for the $10$-$100$ GHz range. 
%\cite{Tosi2012,Wertz2012}

The paper is organized as follows. Section II introduces the concept of a polariton cascade phonon laser based on a wedged stripe. A theoretical description of the device energy levels based on effective models is then included in Section III. Section IV presents the spectroscopic and time-resolved experimental evidence of polariton induced phonon lasing under continuous wave optical pumping, namely i) the observation of inter-level locking,  ii) the emergence of equidistant sidebands, and iii) the observation of periodic oscillations in the time-delayed autocorrelation function $g^{(1)}$. %Section V introduces the modeling of the reported polariton cascade phonon lasing. 
Finally, conclusions and future prospects are drawn in Section V.
% VI. 
%%%%%%%%%%%%%%%%%%%%%%%%%%%%%%%%%%%%%%%%%%%%%
\begin{figure*}[t]
 \begin{center}
    \includegraphics[trim = 0mm 0mm 0mm 0mm,clip=true, keepaspectratio=true, width=0.98 \linewidth,angle=0]{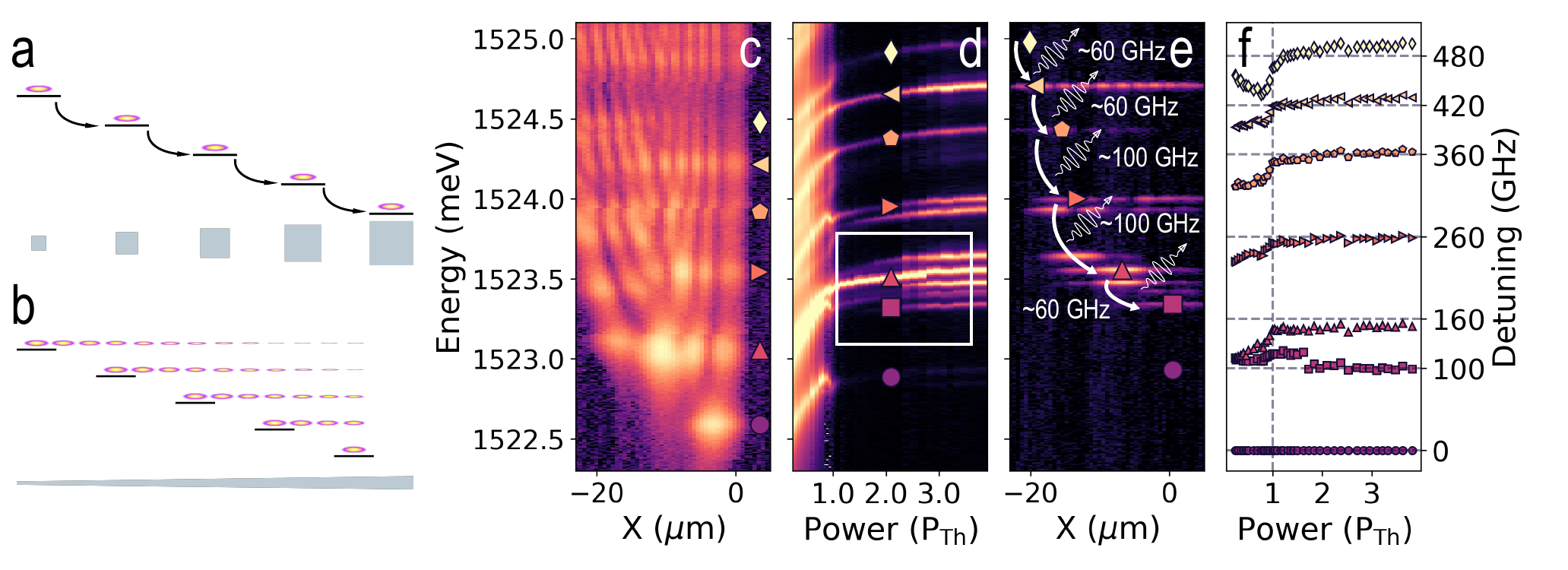}
\end{center}
\vspace{-0.8 cm}
%\hspace{-1.6 cm}
\caption{\textbf{A graded quantum cascade phonon laser.} 
\textbf{(a)} Scheme of a quantum cascade phonon laser (QCPL) based on discrete localized states with energy tuned by the polariton trap size. 
\textbf{(b)} Scheme of a QCPL based on a graded stripe. Note the extended nature of the polariton states, leading to much larger inter-level overlap.
\textbf{(c)} Color map of energy and spatially resolved spectra obtained for low laser powers (below condensation threshold $P \ll P_\mathrm{th}$) with non-resonant cw excitation using a $\sim 5~\mu$m diameter spot localized at $\sim 8~\mu$m from the larger end of the wedged stripe, set at $X=0$.
\textbf{(d)} Power dependence of the spatially-integrated spectra similar to that in \textbf{(c)}. The white dashed rectangle indicates the region detailed in Fig.~\ref{Fig3}(a).
\textbf{(e)} Color map of energy and spatially resolved spectra obtained for large excitation power ($P \gg P_\mathrm{th}$). Arrows indicate phonon-induced transitions. Note how they evolve from left to right following the lateral-confinement induced shift of polariton states. 
\textbf{(f)} Energy levels derived from \textbf{(d)} for the orbital states identified in all panels with the same symbols. All values are given respect to the lower ground state (solid circles). Note that at $P \sim P_\mathrm{Th}$ (indicated with a vertical dashed line) the levels suddenly jump and at higher powers stabilize at separations that are approximately stable and close to $\sim 60$ and $\sim 100$~GHz (dashed horizontal lines).
}
\label{Fig1}
\end{figure*}
%%%%%%%%%%%%%%%%%%%%%%%%%%%%%%%%%%%%%%%%%%%%%

%%%%%%%%%%%%%%%%%%%%%%%%%%%%%%%%%%%%%%%%%%%%%%%%%%%%
\section{Concept of a polariton condensate cascade phonon laser in a wedged stripe\label{Macroladder}}
%%%%%%%%%%%%%%%%%%%%%%%%%%%%%%%%%%%%%%%%%%%%%%%%%%%%
The concept of a QCPL is schematized in Fig.~\ref{Fig1}a with a line of  square polariton traps of varying size connected through tunneling. Due to the reduction of the lateral confinement the energy levels decrease in energy from left to right. A polariton condensate induced in the left-most trap will cascade down the ladder of states, emitting phonons in its path if the trap sizes (and hence the energy levels separation) are appropriately engineered to attain resonant conditions. In principle, the ladder of resonant energy levels could be set by construction and, on operation, be finely tuned using non-uniform laser illumination and the related interaction-induced blue-shift of the polariton states~\cite{Wertz2010}. Such a design, while conceptually parallel to that of a QCL, has however some weaknesses. First, the reduced inter-trap overlap of the polariton wavefunctions limits the magnitude of the corresponding optomechanical coupling~\cite{Chafatinos2020,Reynoso2022}. Moreover, phonons also become confined in the traps~\cite{Fainstein2013,Anguiano2017,Chafatinos2023,Polaromechanics2023}, and thus are limited to contribute collectively to final state stimulation in the device. As a way to solve these two issues, and conserve a ladder-like structure of energy states, we propose a continuous system (a wire) of varying lateral thickness as schematized in Fig.~\ref{Fig1}b. As will be shown later, polariton levels form a ladder of states of decreasing energy from the thinner towards the wider edge. Unlike the coupled traps, these states spatially extend from the thicker side of the stripe to varying positions towards the thinner edge, the latter positions being determined by the lateral confinement. Based on the parallel between photon and phonon confinement in these cavities, the same character of the modes is expected for the device acoustic vibrations, which will thus affect non-locally many different polariton levels.

The studied system is based on $\mu$m-sized structures created by micro-structuring the spacer of an (Al,Ga)As microcavity in-between growth steps by molecular beam epitaxy. 
The sample consisted of a top(bottom) distributed Bragg reflector (DBR) formed by $25(33) \lambda/4,\lambda/4$ Al$_{0.15}$Ga$_{0.85}$As/Al$_{0.90}$Ga$_{0.10}$As layers, embedding a $3\lambda/2$ cavity with four $15$~nm GaAs quantum wells (QWs). Etching of the microcavity spacer prior to the growth of the top DBR into regions of smaller and larger thickness gives rise to cavity modes of higher or lower energy, respectively~\cite{Kuznetsov2018}. These provide the means to define optical and phonon wells and barriers in spatially tailored effective potentials~\cite{Chafatinos2023}. The etching is performed far from the QWs embedded in the spacer layer, so that the high quality of the resulting excitonic system remains preserved. To display strong optomechanical phenomena, the QWs are slightly displaced from the position of the maximum of the cavity optical field $E$, so that the product $sE^2$, with $s$ the confined phonon strain, is maximized at the position of the QWs~\cite{Chafatinos2023}. The DBRs confine the photons and phonons in the $z$ (growth) direction~\cite{Fainstein2013}, while the patterning and etching performed in the spacer determines the lateral confinement~\cite{Chafatinos2023}. 

Figures~\ref{Fig1}(c-f) summarize the main results on a series of experiments performed on a specific structure that is $80 \mu$m long, varies its lateral dimension (from left to right) from $0.5 \mu$m to $2 \mu$m, and which was optically pumped at a distance of $\sim 8 \mu$m from the wider side (defined as $X=0$). The experiments were performed at $5$~K, and the excitation was done non-resonantly ($\sim 1.59$~eV) well above the confined polariton levels ($\sim 1.52$~eV) using a single-mode stabilized Matisse (Spectra Physics) continuous wave laser. Non-resonant excitation leads to the formation of an exciton-reservoir, with a lateral extension set by the laser pump diameter (in our case, $\sim 5 \mu$m), that feeds the polariton trap leading to the detected light emission. Figure~\ref{Fig1}c shows the spatially resolved photoemission spectra obtained at low excitation powers, which is characterized by an almost continuum of states starting at the lower edge close to $\sim1522.5$~meV. On increasing the excitation power, as presented in Fig.~\ref{Fig1}d, the levels first blue-shift and then, after a certain power, a subgroup of them becomes brighter and narrower. The blue-shift is a consequence of polariton-polariton, and polariton-reservoir interactions. The non-linear increase in intensity and line narrowing are typically identified as a signature of condensation of the polaritons. Initially one of the states located around $1523.5$~eV captures most of the intensity (at the threshold power $P_\mathrm{th} \sim 15$mW), and then around $2 P_\mathrm{th}$ the polariton emission 
noticeably becomes more evenly distributed in a set of states for which the corresponding spatial and spectral distribution is presented in Fig.~\ref{Fig1}e. It follows from this latter figure that the levels decrease in energy, with their center of mass going from left (smaller lateral size) to right (larger lateral size of the stripe), as expected. Moreover, the states are spectrally separated in clusters, some of which have an internal structure with a fan-like distribution (e.g., the one highlighted with a dashed rectangle in Fig.~\ref{Fig1}d). This internal structure will be discussed in detail in following sections, we start here by analyzing the main component of each set. 

The power dependence of the energy separation between these brighter states, referenced to that of the lowest energy  to correct for the overall interaction-induced blue-shift, is shown in Fig.~\ref{Fig1}f. Though as observed in Fig.~\ref{Fig1}d the states continue to blue-shift up to the highest excitation power, the inter-level energy separation becomes rather flat. Most notably, the energy separation of all the selected states suddenly jumps when condensation is established, and becomes close to either $60$ or $100$~GHz. These latter values coincide with the first two overtones of the confined phonon mode of the planar cavity: fundamental mode at $\nu_\mathrm{m}^{(0)} \sim 20$~GHz, and overtones at $\nu_\mathrm{m}^{(n)}=(2n+1)\nu_\mathrm{m}^{(0)}$, with $n=1, 2, ...$~\cite{Fainstein2013}. As previously demonstrated~\cite{Chafatinos2023,Ramos2024} such asynchronous locking is a manifestation of mechanical coherent oscillation at these phonon frequencies. The results in Fig.~\ref{Fig1}f indicate that this occurs concomitantly with polariton condensation. As we will demonstrate in what follows, altogether the experiments in Fig.~\ref{Fig1} are compelling evidence of polariton condensates cascading down a ladder of states, accompanied by the driving of multifrequency coherent mechanical vibrations. 

%%%%%%%%%%%%%%%%%%%%%%%%%%%%%%%%%%%%%%%%%%%%%
\begin{figure*}[t]
    \centering 
\includegraphics[trim = 0mm 0mm 0mm 0mm, clip=true, keepaspectratio=true, width=1.8\columnwidth,angle=0]{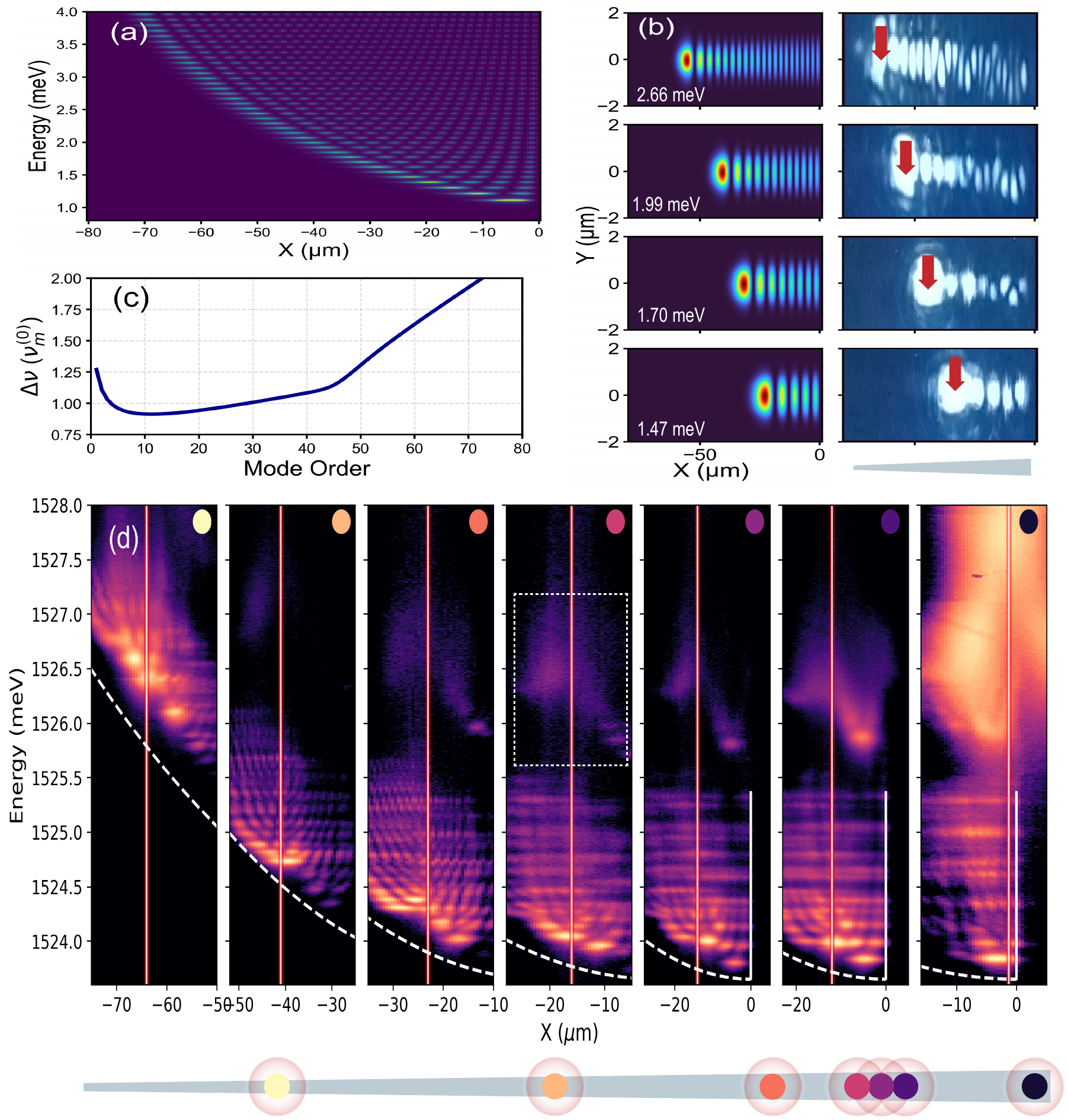}
 \caption{
\textbf{Ladder of polariton states in a wedged stripe.}
\textbf{(a)} Energy and spatial distribution of the calculated levels. 
\textbf{(b)} Spatial distribution of the polariton modes of increasing energy (from bottom to top). Left is calculated, right is experimental. For the latter the position of excitation is indicated with down red arrows. 
\textbf{(c)} Calculated energy separation between successive modes of the studied stripe, with energies given in units of the confined fundamental vibrational mode ($\nu^{(0)}_\mathrm{m}\sim 20$~GHz), and the modes are labeled with an integer number with increasing energy. For the first $\sim 50$ modes the energies increase close to linear and with a separation $\Delta \nu$ that is close to $\nu^{(0)}_\mathrm{m}$. 
\textbf{(d)} Energy and spatially resolved spectra obtained for low laser powers (below condensation threshold $P \ll P_\mathrm{th}$) as in Fig.~\ref{Fig1}(c), for different spot positions along the graded stripe as indicated in the bottom scheme. The dashed curves indicate the parabolic-like spatial dependence of the effective potential. The dashed rectangle highlights the laser-induced blue-shift of the states of the stripe (this shift is also present for the lower energy modes, but is enhanced for the higher energy states with larger excitonic character).
    }
\label{Fig2}
\end{figure*} 
%%%%%%%%%%%%%%%%%%%%%%%%%%%%%%%%%%%%%%%%%%%%%

%%%%%%%%%%%%%%%%%%%%%%%%%%%%%%%%%%%%%%%%%%% 
 \section{Energy levels of a wedged stripe}
%%%%%%%%%%%%%%%%%%%%%%%%%%%%%%%%%%%%%%%%%%% 
To advance on the interpretation of the experimental results shown in Fig.~\ref{Fig1} we describe next the structure of energy levels of a wedged stripe. Figure \ref{Fig2}(a) shows the calculated polariton energy levels for a $80 \mu$m long stripe, with lateral dimension varying linearly from $0.5 \mu$m to $2 \mu$m. Zero energy in this figure corresponds to the polariton mode of a planar structure (i.e., without lateral confinement). We model this using an effective generalized Gross-Pitaesvki equation (gGPE)~\cite{Wouters2007} as described in detail in Appendix A. 
Some examples of the corresponding calculated  polariton eigenmodes are shown at the left of Fig.~\ref{Fig2}(b). Modes with increasing energy become more extended, and have larger number of nodes along the stripe. The states have a stronger amplitude around the turning point in the high-energy (smaller lateral confinement) side of the parabolic-like effective potential induced by confinement, and spatially extend towards the wider edge. Quite notably, as shown in Fig.~\ref{Fig2}(c), roughly the first $50$ energy levels increase rather linearly with mode-number, with a separation $\Delta \varepsilon/\hbar=\Delta \nu \sim \nu^{(0)}_\mathrm{m} = 20$~GHz (this frequency matching is sensitive to the depth of the potential (exciton photon detuning) see Appendix A).

Experimentally, the full structure of modes can be derived from spatially and spectrally resolved intensity maps obtained at low pumping powers ($P \ll P_\mathrm{th}$), as shown in Fig.~\ref{Fig2}(d). These experiments  correspond to the situation where the excitation is focused at different positions along the stripe (indicated below the figure). Notice the resemblance with the energy levels expected from the theory. The spatial distribution of some selected modes is shown in more detail in the right panel of Fig~\ref{Fig2}(c)---the corresponding energy increases from bottom to top. These different experimental examples were obtained by excitation above polariton condensation threshold, and varying the position of the excitation (indicated with down red arrows) to selectively induce situations where mainly a single mode contributes to the emission. As will be shown below, in intermediate positions several modes coexist. Note the highly extended character of the polariton emission and the presence of the characteristic spatial interference fringes, showing that the polariton condensate propagates and conserves coherence throughout the full length of the stripe.

%%%%%%%%%%%%%%%%%%%%%%%%%%%%%%%%%%%%%%%%%%%%%
\begin{figure*}[ht]
    \centering 
    \includegraphics[trim = 0mm 0mm 0mm 0mm,clip=true, keepaspectratio=true, width=2\columnwidth,angle=0]{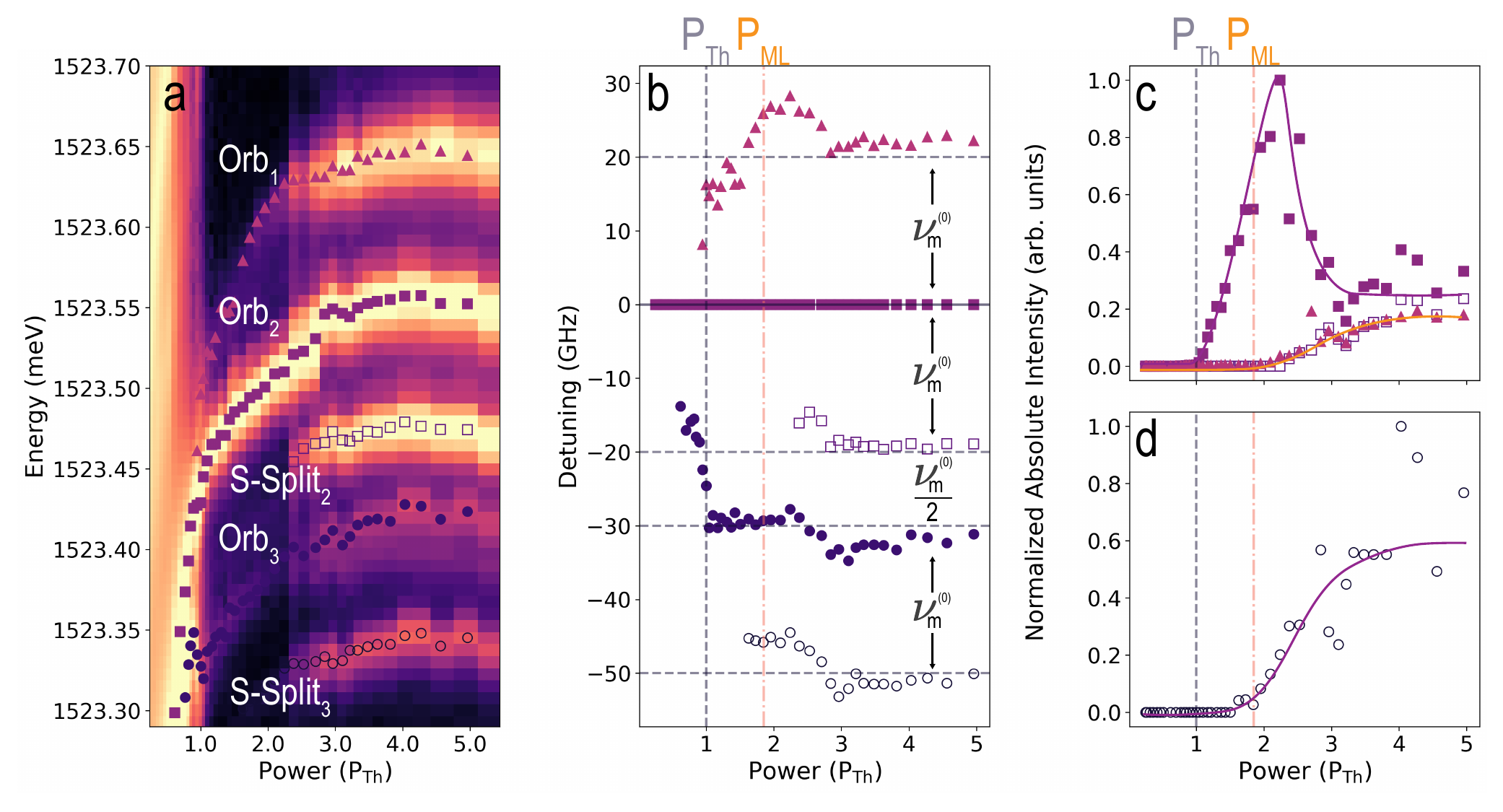}
    \caption{
    \textbf{Mechanically induced ladder of states.} 
    \textbf{(a)} Detail of the power dependence of modes identified with a white dotted rectangle in Fig.~\ref{Fig1}(d). Three distinctive orbital states (Orb) and two spin-split states (S-split) can be identified based on the spatial and polarization properties. The corresponding energies are given in \textbf{(b)}, referenced to the most intense (Orb$_2$) state (to discount the overall interaction-induced blue-shift). Note that the spin-split levels stabilize at one phonon quanta below the corresponding orbital state. And also that Orb$_1$ does it at one phonon-quanta above the most intense Orb$_2$ state.  The corresponding integrated intensity of the peaks are given in \textbf{(c)} and \textbf{(d)}, with peaks identified by the same symbols as in \textbf{(a-b)}. Note that all states besides Orb$_2$ increase their intensity and lock their energies around the same power ($P_\mathrm{ML} \sim 2P_\mathrm{th}$). 
}
    \label{Fig3}
\end{figure*} 
%%%%%%%%%%%%%%%%%%%%%%%%%%%%%%%%%%%%%%%%%%%%%

The laser spot perturbs the effective potential sensed by the trapped polaritons. The focused excitation (position identified with vertical continuous lines on the experiments shown in Fig.~\ref{Fig2}(d)) leads, through Coulomb interactions, to a local blue shift that defines a barrier. This is evident in the experiments for all states but, in particular, for those at higher energy---highlighted with a white dotted rectangle for the spot located at the middle of the wedged stripe in Fig.~\ref{Fig2}(d)---since they have a more excitonic character. This perturbation modifies the trapped states leading to characteristic changes that, however, do not have a significant influence on the polariton cascades discussed in this work. A theoretical and an experimental analysis of the effects on the polariton modes, associated to the presence of the laser pump, are presented in Appendix A and B, respectively.

%%%%%%%%%%%%%%%%%%%%%%%%%%%%%%%%%%%%%%%%%%%%%
\begin{figure*}[ht]
    \centering 
    \includegraphics[trim = 0mm 0mm 0mm 0mm,clip=true, keepaspectratio=true, width=2\columnwidth,angle=0]{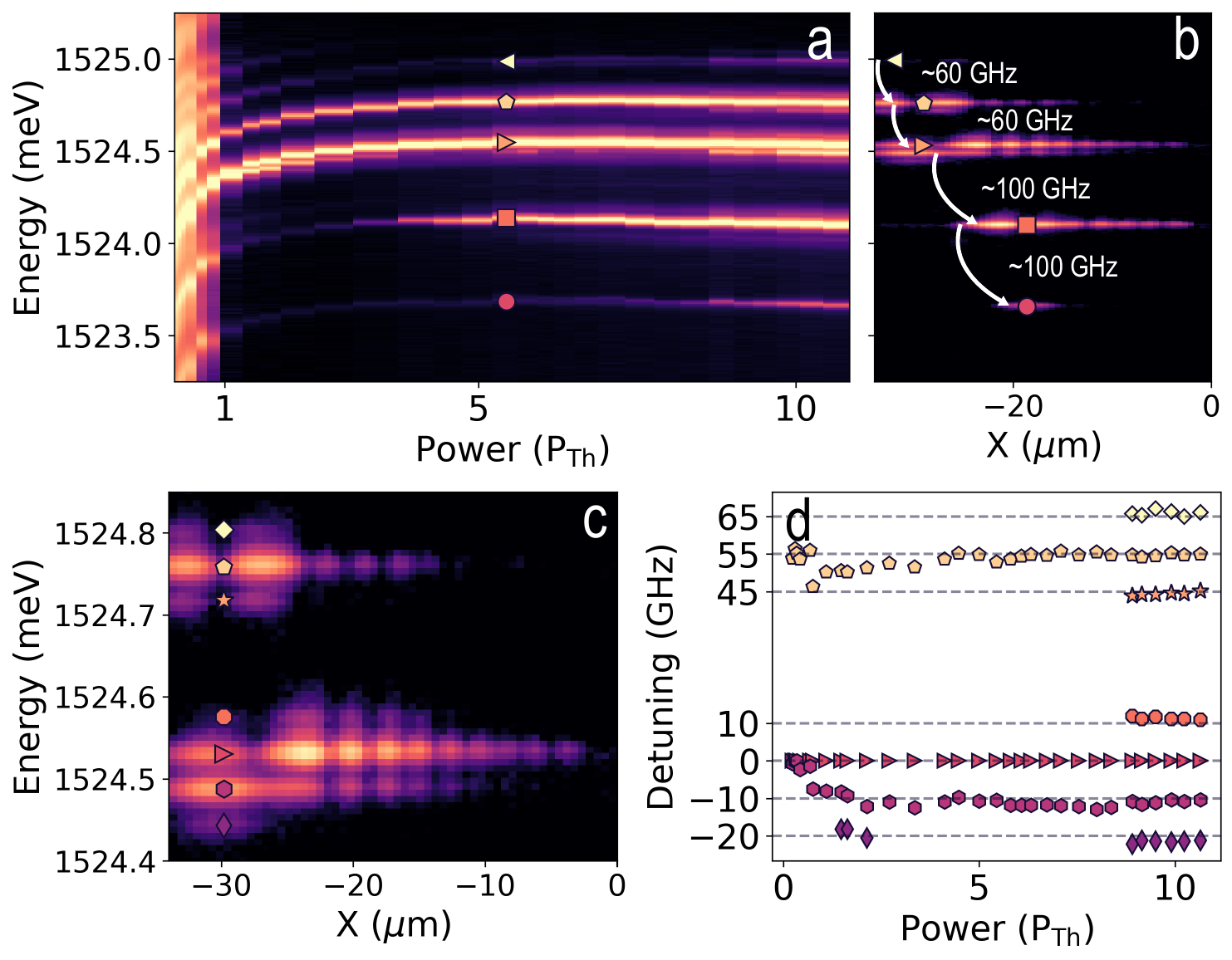}
    \caption{
    \textbf{Mechanically induced sidebands.} 
\textbf{(a)} Power dependence of the spatially-integrated spectra for a laser spot excitation further shifted to the narrow region of the stripe (third from the left in Fig.~\ref{Fig2}(a))
\textbf{(b)} Color map of energy and spatially resolved spectra obtained for large excitation power ($P \gg P_\mathrm{th}$). Note how the ladder of states evolves from left to right following the lateral-confinement induced blue-shift of polariton states. 
Arrows indicate the polariton cascade accompanied by emission of phonons, again as in Fig.~\ref{Fig1}(c) with energy separation given by $\sim 60$ and $\sim 100$~GHz.  
A detail of the spatial distribution of some states is presented in \textbf{(c)} to highlight the emergence of equidistant sidebands. The power dependence of these energies (given respect to the main peak), is presented in \textbf{(d)}. Note that dashed lines in \textbf{(d)} are separated by $\nu^{(0)}_\mathrm{m}/2  \sim 10$~GHz, an indication of period-doubling. 
}
    \label{Fig4}
\end{figure*} 
%%%%%%%%%%%%%%%%%%%%%%%%%%%%%%%%%%%%%%%%%%%%%
%%%%%%%%%%%%%%%%%%%%%%%%%%%%%%%%%%%%%%%%%%%%%%%%%%%%%%%%%%%%%%%%%%%%%%%%%%%%%%%%%%%
\section{Spectral and temporal characteristics of polariton cascade phonon lasing}
%%%%%%%%%%%%%%%%%%%%%%%%%%%%%%%%%%%%%%%%%%%%%%%%%%%%%%%%%%%%%%%%%%%%%%%%%%%%%%%%%%
%%%%%%%%%%%%%%%%%%%%%%%%%%%%%%%%%%%%%%%%%%%%%%%%%%%%%%%%%%
\subsection{Locking of orbital and spin-split polariton states by a mechanical coherent oscillation}
%%%%%%%%%%%%%%%%%%%%%%%%%%%%%%%%%%%%%%%%%%%%%%%%%%%%%%%%%%
As described in the previous section, the wedged wire provides a closely packed distribution of approximately evenly spaced energy levels. The experiments in Fig.~\ref{Fig1}, in turn, demonstrate that only a subset of these states becomes occupied for powers above the condensation threshold. Notably, as illustrated in Fig.~\ref{Fig1}(f), the energy separation between such clusters approaches, with increasing power, the energy of phonon modes vertically confined in the resonator (i.e., $\sim 60$~GHz and $\sim 100$~GHz for the shown case). It has been previously demonstrated that such asynchronous locking stems from the presence of a harmonic time modulation of the Josephson coupling between polariton modes~\cite{Chafatinos2023}, the physics of which is linked to synchronization~\cite{Ramos2024}. In this case, the origin of the time modulation is understood to be due to the presence of coherent mechanical oscillations induced by the polaritons themselves~\cite{Chafatinos2020,Chafatinos2023}. The magnitude of the locked splitting is evidence by itself of the vibrational frequencies involved in the coherent time modulation. The observation of such a strong effect is a consequence of the exciton-enhanced optomechanical coupling present in this system~\cite{Zambon2022,Sesin2023,Polaromechanics2023}, and is facilitated by polariton nonlinearities and dissipation~\cite{Chafatinos2023,Ramos2024}. 
%As shown in Figs.~\ref{Fig1}(d) and (f), the states lock at these energy differences ($\sim 60$ and $\sim 100$~GHz), and when this locking occurs, it is accompanied by a redistribution of spectral weight between clusters. 
As highlighted with the white dotted square in Fig.~\ref{Fig1}(d), this locking is also accompanied, at even higher powers, by the development of internal structure within clusters. This is further discussed in the following and is detailed in Fig.~\ref{Fig3}. 

From the spatial distribution of the modes, whose power dependence is presented in Fig.~\ref{Fig3}(a), we identify three states with different orbital character. One of them, labeled Orb$_2$, is initially the most intense while the other two orbital states are dimmer until $P \gtrsim 2 P_\mathrm{th}$ where a sudden increase of the relative intensity occurs [see Fig.~\ref{Fig3}(c)]. Interestingly, together with this change of relative amplitude, two of the orbital states split in two. The doubling is known to reflect the pseudospin degree of freedom of the polariton condensates~\cite{Shelykh2005,Ohadi2015,Gnusov2020,Carraro2024}. Spin-split states are identified through the polarization of the emitted light, and because they share the same spatial distribution as the ``parent'' orbital states (in Fig.~\ref{Fig3}(a), for example, S-Split$_2$ identifies a spin-split state that has the same spatial distribution as the orbital state Orb$_2$). The power dependence of the energy separation between these modes, referred to the most intense one (Orb$_2$), is presented in Fig.~\ref{Fig3}(b). Similar to the ``macro'' set of brighter modes described in Fig.~\ref{Fig1}, this ``micro'' perspective shows that the levels also asynchronously lock their energy separation when they become brighter. Moreover, this energy separation is not arbitrary. The two spin-split states are red shifted quite precisely by the mechanical fundamental frequency $\nu^{(0)}_\mathrm{m} \sim 20$~GHz, the upper Orb$_1$ state is blue-shifted by the same amount respect to the most intense Orb$_2$, and the state Orb$_3$ is separated by approximately $\nu^{(0)}_\mathrm{m}/2$ from the Orb$_2$ doublet. We note that the sudden pseudospin splitting with increasing excitation power in polariton condensates has been identified as a spontaneous breaking of time translation symmetry, leading to a time-crystal behavior involving a stable Larmor-like precession of the polariton pseudospins. The fact that the splitting amounts to $\nu^{(0)}_\mathrm{m} \sim 20$~GHz is evidence that these stable limit-cycles drive confined coherent phonons that, in turn, lock and stabilize the pseudospin precession~\cite{Carraro2024}. The asynchronous locking of orbital states at frequencies corresponding to integer numbers and fractions of $\nu^{(0)}_\mathrm{m} \sim 20$~GHz, is also a consequence of the presence of such self-driven coherent phonons modulating the polariton modes, and of the non-linearities that characterize this polariton system~\cite{Chafatinos2023,Ramos2024}.

The corresponding power dependence of the intensity of the lines is presented in Figs.~\ref{Fig3}(c) and \ref{Fig3}(d). The same symbols are used in all four panels of Fig.~\ref{Fig3} to identify the observed modes. The intensities are normalized to the maximum in each panel. The most intense orbital state Orb$_2$ is macroscopically occupied and increases its intensity above condensation threshold $P_\mathrm{th}$. At a higher power, it decreases and distributes its weight with the other orbital and spin-split states. We identify the power at which the additional lines get brighter, with a threshold-like behavior, as $P_\mathrm{ML}$ (ML stands for ``mechanical lasing''). We identify this second threshold $P_\mathrm{ML}$ with the establishment of self-induced mechanical lasing at $\nu^{(0)}_\mathrm{m} \sim 20$~GHz. This is in addition to that at the higher overtones $\nu_m^{(1)}$ and $\nu_m^{(2)}$ established at condensation threshold and evidenced by the locking of the macro-structure of levels discussed in Sect.~\ref{Macroladder}. While the presented experimental data point to the existence of two separate phonon lasing thresholds, we note that the second one at $P_\mathrm{ML} \sim 2 P_\mathrm{th}$ leads to an enhanced redistribution of populations between all macroscopically occupied states, as is clear in Fig.~\ref{Fig1}(d). We understand this spreading of intensities between the micro and macro sets as an indication of non-linearities coupling all the lasing phonon modes, as will be further discussed below when presenting the time-delayed autocorrelation function ($g^{(1)}(\tau)$) measurements.
%%%%%%%%%%%%%%%%%%%%%%%%%%%%%%%%%%%%%%%%%%%%%%%%%%%%%%%%%%%%
\subsection{Mechanically induced sidebands and period doubling}
%%%%%%%%%%%%%%%%%%%%%%%%%%%%%%%%%%%%%%%%%%%%%%%%%%%%%%%%%%%% 
%%%%%%%%%%%%%%%%%%%%%%%%%%%%%%%%%%%%%%%%%%%%%
\begin{figure*}[!hht]
 \begin{center}
    \includegraphics[trim = 0mm 0mm 0mm 0mm,clip=true, keepaspectratio=true, width=1.8 \columnwidth,angle=0]{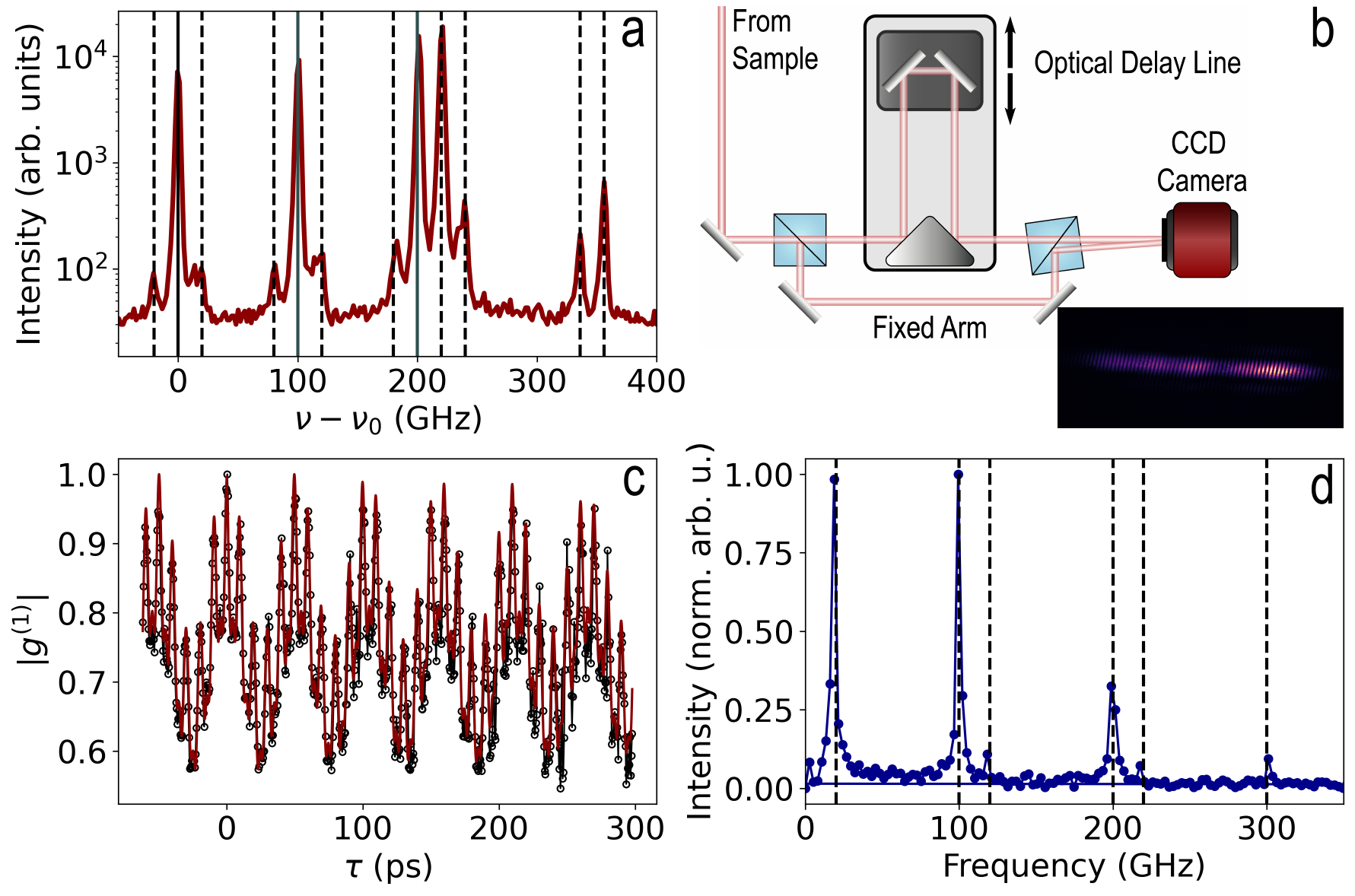}
\end{center}
\vspace{-0.8 cm}
%\hspace{-1.6 cm}
\caption{\textbf{Time-resolved autocorrelation function $g^{(1)}(\tau)$.} 
Measurement performed for a third spot position different from that leading to the experiments in Figs.~\ref{Fig1} and \ref{Fig3}. The corresponding spectrum is shown in \textbf{(a)}, with energies referred to that of the main lower energy peak. Vertical dashed lines are separated by 20~GHz. A scheme of a $g^{(1)}(\tau)$ measurement is illustrated in \textbf{(b)}. A detail of the first 300~ps of $g^{(1)}(\tau)$ is presented in \textbf{(c)}. \textbf{(d)} The Fourier transform of the $g^{(1)}(\tau)$ measurement in \textbf{c} is given in  \textbf{(d)}. Vertical dashed lines correspond to $20, 100, 120, 200, 220$ and 300~GHz.
}
\label{Fig5}
\end{figure*}
%%%%%%%%%%%%%%%%%%%%%%%%%%%%%%%%%%%%%%%%%%%%%

Besides the locking of orbital states and the splitting of pseudo-spin degeneracies, both to values that match the phonon frequencies, the stimulated emission of coherent phonons is manifested by the emergence of equally spaced sidebands~\cite{Chafatinos2020}. These sidebands are much weaker than the main lines, and thus difficult to observe, but in some cases become noticeable. One such a case observed in our platform is illustrated in Fig.~\ref{Fig4}, where we present a power dependence study of the spectra, similar to that previously discussed but performed on a different position along the wedged stripe. As before, a reduced set of polariton modes becomes strongly populated  as function of the excitation power (Fig.~\ref{Fig4}(a)). Furthermore, as discussed in the previous sections, these modes are spatially distributed like a ladder descending from the thinner to the wider side of the stripe. Conspicuously, they are separated either by $\sim 60$~GHz or $\sim 100$~GHz (Fig.~\ref{Fig4}(b)), the frequencies of confined phonon vibrations of the structure. A close view on the high-power spatial map evidences that some of these levels present side-bands (see Fig.~\ref{Fig4}(c)). The corresponding power dependence of the peak positions are displayed in Figs.~\ref{Fig4}(d). Interestingly,  the sidebands are separated by $\nu_m^0/2 \sim 10$~GHz. As shown in Ref.~\cite{Carraro2024}, this can be interpreted as evidence of the presence of phonon lasing, leading to a mechanical modulation of the polariton condensates at $\nu^{(0)}_\mathrm{m} \sim 20$~GHz. This modulation results in limit-cycles displaying, for the case shown, another consequence of time crystalline behavior, namely period-doubling~\cite{Ramos2024,Carraro2024}. 
%%%%%%%%%%%%%%%%%%%%%%%%%%%%%%%%%%%%%%%%%%%%%%%%%%%%%%%%%%%%%%%%%%%%%%%
 \subsection{Time-delayed autocorrelation function $g^{(1)}(\tau)$}
%%%%%%%%%%%%%%%%%%%%%%%%%%%%%%%%%%%%%%%%%%%%%%%%%%%%%%%%%%%%%%%%%%%%%
The observation of both asynchronous-locking and equidistance sidebands of the polariton condensate modes  are supporting evidence for the emergence of internal dynamics involving and modulating the exciton-polaritons. That is, the emergence of coherent mechanical oscillations being established by the polariton bosonic cascade~\cite{Chafatinos2020,Chafatinos2023,Ramos2024}. In addition to this, one can obtain more direct time-domain evidence of the self-established vibrational dynamics in an experiment with continuous wave excitation by measuring the time-resolved spatial first-order coherence function $g^{(1)}(\tau)$~\cite{Baryshev2022}. To that end, we acquire simultaneously the high-resolution spectrum and the autocorrelation function $g^{(1)}(\tau)$ for an excitation of the wedged stripe in a similar condition to that illustrated in Figs.~\ref{Fig1} and \ref{Fig3}. The obtained spectrum is displayed in Fig.~\ref{Fig5}(a). It contains all the features described so far which are associated to phonon lasing, namely: i) the selective macroscopic occupation of a sub-set of polariton levels asynchronously locked and separated by $\nu_\mathrm{m}^{(2)} \sim 100$~GHz; ii) the splitting of pseudospin states also locked at a separation of $\sim 20$~GHz; and, finally, iii) the emergence of symmetric sidebands, in this case separated by $\nu_\mathrm{m}^{(0)} \sim 20$~GHz~\cite{Chafatinos2020}. A simplified scheme of the $g^{(1)}(\tau)$ experiment is presented in Fig.~\ref{Fig5}(b), together with an example of the obtained auto-correlation interference pattern derived by focusing and interfering together both the directly collected light (fixed arm path) and its time-delayed image (delay line path). The measured $g^{(1)}(\tau)$ is presented in Fig.~\ref{Fig5}(c). It evidences the existence of both fast and relative slow periodic oscillations, evolving almost without decay in the sampled $\sim 300$~ps time-window. The corresponding Fourier transform is shown in Fig.~\ref{Fig5}(d). Clear strong frequency components at 
$\nu_\mathrm{m}^{(0)} \sim 20$~GHz and $\nu_\mathrm{m}^{(2)} \sim 100$~GHz are evident, plus multiples and combinations of them: $120$~GHz, $200$~GHz, $220$~GHz, and $300$~GHz. We interpret this latter observation as providing additional evidence of phonon non-linearities in the presence of multi-wavelength phonon lasing. 
%%%%%%%%%%%%%%%%%%%%%%%%%%%%%%%%%%%%%%%%%%%%%%%%%%%%%%%%%%%%%%%%%%%
 \subsection{On the efficiency of bosonic cascade phonon lasers}
%%%%%%%%%%%%%%%%%%%%%%%%%%%%%%%%%%%%%%%%%%%%%%%%%%%%%%%%%%%%%%%%%%
It has been argued that the quantum efficiency of bosonic cascade lasers can be several orders of magnitude larger than the corresponding fermionic cascades. In fact, lasing with high quantum efficiency, above unity, was theoretically shown for the THz bosonic cascade lasers in Ref.~\cite{Liew2013}. This is in fact also the case of the bosonic cascade phonon lasers demonstrated here. The non-resonant laser excitation is done through the DBR's stop-band edge, where the reflectivity is close to zero. Calculation of the absorption for the specific case of our microcavity with four embedded QWs, indicates that approximately $50\%$ of the photons are converted to e-h pairs. Non-radiative losses in these materials are negligible. Thus, one can safely assume that one out of every two excitation photons generates e-h pairs, and that all of these e-h pairs are converted to polaritons and finally emitted as photons leading to the condensate spectra as shown in Figs.~\ref{Fig1} and \ref{Fig4}. Experimentally, we can make a gross estimation of the number of phonons emitted, simply by integrating the relative intensity of the corresponding cascade ``clusters'', and counting the number of phonons required to arrive at the respective polariton levels, assuming that the cascade initiates at the higher detected macroscopically occupied polariton state (for example, the one labeled with a rhomb in Fig.~\ref{Fig1}(c)). As shown in the examples discussed above, all the presented cascades are similar but differ in the details, and can be rather finely tuned by displacing the excitation laser spot along different positions on the wedged stripe.  The quantum efficiency estimated in this way for the cascade spectra shown in Figs.~\ref{Fig1}(e), \ref{Fig4}(b) and \ref{Fig5}(a), can be as large as $1$ for either the $60$~GHz or the $100$~GHz confined vibration. Here we define the quantum efficiency as the number of phonons emitted per exciting photon. 
%%%%%%%%%%%%%%%%%%%%%%%%%%%%%%%%%%%%%%%%%%%%%%%%%%%%%%%%%%%%%%%%%%%%%%%%%%% 
\section{Simplified model of phonon emission}
%%%%%%%%%%%%%%%%%%%%%%%%%%%%%%%%%%%%%%%%%%%%%%%%%%%%%%%%%%%%%%%%%%%%%%%%%%%  
There are several possibilities to model the effect of the polariton-phonon coupling. Some of them were successfully used to describe the emergence of stimulated emission of phonons and phonon induced locking on different simpler (two-level) systems~\cite{Chafatinos2023,Ramos2024,Reynoso2022,Carraro2024}. While the results presented in the previous sections share some common features with those, a simple model able to describe the overall phenomenology of the observed different phonon cascades (e.g., Figs. \ref{Fig1}, \ref{Fig3} and \ref{Fig4}), involving a multi-level and multi-phonon system, is still lacking. 

Nevertheless, we introduce here a simplified effective model that shows some of the key features, that might serve as a guidance for future work. 
Based on the analysis of the generalized Gross-Pitaesvkii equation  for the lower polariton field \cite{Wouters2007} presented in Appendix \ref{ap:gGPE}, we consider a set $\{\phi_j\}$ of $N_p$ polariton modes with energies $\{\varepsilon_j\}$ corresponding to the eigenmodes of the wedged trap which are coupled through a phonon mode of the cavity of energy $\hbar\Omega$. The  adimensional generalized coordinate of the latter is denoted $q=b+b^{\dagger}$ with $b^{\dagger}$ ($b$) the creation (annihilation) operator of such phonon mode. The corresponding equations of motion are then given by~\cite{Reynoso2022}
%%%%%%%%%%%%%%%%%%%%%%%%%%%%%%%%%%%%%%%%%%%%%%%%%%%%%%%%%%
\begin{eqnarray}
\nonumber
i\hbar\dot{\phi}_j(t) &=& \left(\varepsilon_j+\frac{i\hbar \gamma}{2}(n_j-1)\right)\phi_j(t)+q(t)\sum_{k=1}^{N_p} \hbar J_{jk}\phi_k(t)\,,\\
\nonumber
\ddot{q}(t)&=&-\Omega^2 q(t)-\Gamma\dot{q}(t)-4\Omega\rho_0\mathrm{Re}\left(\sum_{k=1}^{N_p} J_{jk}\phi_j^*(t)\phi_k(t)\right)\,.\\
\label{eq:general}
\end{eqnarray}
%%%%%%%%%%%%%%%%%%%%%%%%%%%%%%%%%%%%%%%%%%%%%%%%%%%%%%%%%%
Here, $n_j=p_j/(1+|\phi_j|^2)$ represents the gain from the exciton reservoir, $p_j$ being the power of the pump laser in units of the threshold power, $\rho_0$ a reference polariton number used to defined the normalized amplitudes $|\phi_j|^2$ \cite{Chafatinos2023}, $\gamma$ is the lower polariton decay rate, $\hbar J_{jk}$ the polariton-phonon coupling constant and $\Gamma$ the damping rate of the phonon. For simplicity we ignore here the polariton-polariton interaction (assumed to be small) and consider that the interaction with an inactive reservoir gives an overall shift for the considered modes included on the mode's energy. We solve these equations assuming that a single mode is pumped, say $j=0$, so that $p_j=p \delta_{j0}$.

To grasp the main features of this model, let us first analyze the case of  only two polariton modes separated by an energy $\hbar\Omega$ (`resonant case'). Making a transformation to a rotating frame, $\phi_j(t)\rightarrow \bar{\phi}_j e^{-i \omega_j t}$,  with $\omega_j=\varepsilon_j/\hbar$, we have
%%%%%%%%%%%%%%%%%%%%%%%%%%%%%%
\bea
\nonumber
i\dot{\bar{\phi}}_0 &=&J_{01} q(t) \bar{\phi}_{1}e^{-i\Delta\omega t}+i(\gamma/2)(n_0-1) \bar{\phi}_{0}\,,\\
\label{EOM2x2}
i\dot{\bar{\phi}}_1 &=&-i(\gamma/2)\bar{\phi}_{1}+J_{01} q(t) \bar{\phi}_{0}e^{i\Delta\omega t}\,,\\
\nonumber
\Ddot{q}(t)&=&-\Omega^2q(t)-\Gamma\Dot{q}(t)-4\Omega \rho_0J_{01}\mathrm{Re}(\bar{\phi}_{0}\bar{\phi}_{1}^*e^{i\Delta\omega t})\,.
\eea
%%%%%%%%%%%%%%%%%%%%%%%%%%%%%%
with $\Delta\omega=\omega_1-\omega_0=s\Omega$ and $s=\pm1$ depending on where $\varepsilon_1$ is above or below $\varepsilon_0$. Here we took $J_{jj}=0$ as only the coupling between modes can lead to the generation of phonons in this model. If we now assume that the self consistent amplitude of the phonon field will be of the form $q(t)=2A \cos(-s\Omega t+\theta)$  and that $|J_{01}A/\Delta\omega|\ll 1$ we can make a rotating wave approximation and keep only the time independent terms so that $\bar{\phi}_j$ admit a stationary solution. In that case, from the last two equations in Eq.~\eqref{EOM2x2} we obtain 
%%%%%%%%%%%%%%%%%%%%%%%%%%%%%%
\bea
\nonumber
\bar{\phi}_i &=&-i\frac{2J_{01}Ae^{i\theta}}{\gamma}\bar{\phi}_0\,,\\
s \Gamma  A &=&-\frac{4J_{01}^2\rho_0}{\gamma}A|\bar{\phi}_0|^2\,.
\label{conditions}
\eea
%%%%%%%%%%%%%%%%%%%%%%%%%%%%%%
From here, it is clear that a self-consistent phonon ($A\neq0$) can only appear if $s=-1$, so that the unpumped mode is below the pumped one, $\varepsilon_1=\varepsilon_0-\hbar\Omega$. This can be seen as the usual case in optomechanics where a `blue'-detuned laser (in this case the role of pumped mode) is needed to induce the oscillations~\cite{AspelmeyerRMP2014}.
We hence obtain that 
%%%%%%%%%%%%%%%%%%%%%%%%%%%%%%%
\begin{equation}
|\bar{\phi}_0|^2=\Gamma\gamma/4J_{01}^2\rho_0\,,\qquad |\bar{\phi}_1|^2=(\Gamma/\gamma\rho_0) A^2\,.    
\end{equation}
%%%%%%%%%%%%%%%%%%%%%%%%%%%%%
Two important conclusions can be draw from here: (i) when the phonon oscillation is induced by the cascade process, the pumped mode saturates so that the cooperativity $C=4J_{01}^2n_0/\Gamma\gamma$, with $n_0=|\bar{\phi}_0|^2\rho_0$, is equal to one; (ii)  the population of the second mode is proportional to the amplitude of the phonon (see discussion in previous section regarding Figs.~\ref{Fig3}(c) and \ref{Fig3}(d)).
The power dependence of the phonon and the second mode can be obtained by replacing Eqs.~\eqref{conditions} into the first equation in Eq.~\eqref{EOM2x2} from where we get that $A^2=(\gamma^2/4J_{01}^2)(p-p_0)/p_0$ with $p_0=1+\Gamma\gamma/4J_{01}^2\rho_0$. Note that the phonon frequency is not present in these equations due to the perfect resonant conditions. A small detuning, smaller than $\gamma$, does not prevent the stimulation of the phonon.  
%%%%%%%%%%%%%%%%%%%%%%%%%%%%%%%%%%%%%%%%%%%%
\begin{figure}[t]
 \begin{center}
    \includegraphics[trim = 0mm 0mm 0mm 0mm,clip=true, keepaspectratio=true, width=0.98\columnwidth,angle=0]{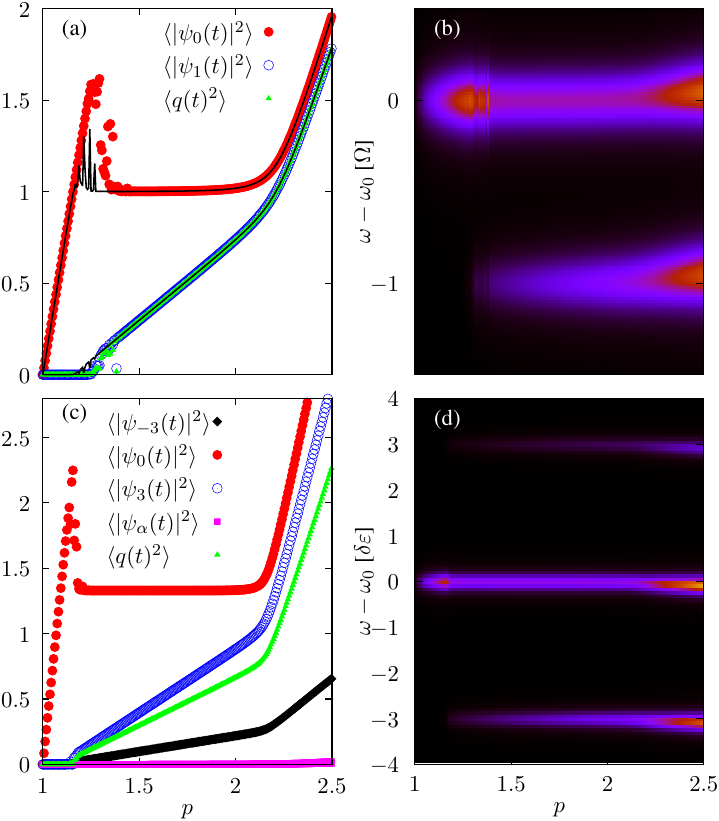}
\end{center}
\vspace{-0.8 cm}
%\hspace{-1.6 cm}
\caption{Time average population of the polariton modes, $\langle|\phi_j(t)|^2\rangle$, and the phonon, $\langle q^2(t)\rangle$, as a function of the laser pump power $p$ obtained from the numerical solution of Eq.~\eqref{eq:general} with $N_p=2$ (a) and $N_p=7$ (c). The polariton modes are normalized to $|\bar{\phi}_0|_{\mathrm{sat}}^2$ and the phonon to $\gamma^2/2J_{01}^2$. In both cases the onset of the phonon emission in signaled by the increase of population of the resonant polariton modes. The solid thin (black) line in (a) shows the corresponding amplitudes for the case of a larger initial amplitude of the phonon. In (c) $|\psi_\alpha(t)|^2$ refers to the polariton modes not explicitly indicated by the labels. Panels (b) and (d) show the corresponding full power spectrum of the polariton modes. Note that in (d) the intermediate polariton modes (which are not resonant) remain empty.}
\label{Fig6}
\end{figure}
%%%%%%%%%%%%%%%%%%%%%%%%%%%%%%%%%%%%%%%%%%%%

Figure \ref{Fig6}(a) shows the asymptotic time average of the amplitude square of each mode, $\langle|\phi_j(t)|^2\rangle$, and of the phonon, $\langle q^2(t)\rangle$, as a function of the power of the laser pump $p$ as obtained from the numerical solution of the full sets of equations in Eq.~\eqref{EOM2x2} with a single(random) initial condition. The polariton and phonon averages are normalized by $|\bar{\phi}_0|_{\mathrm{sat}}^2=\Gamma\gamma/4J_{01}^2\rho_0$ and $\gamma^2/2J_{01}^2$, respectively. Figure \ref{Fig6}(b) shows the corresponding spectrum. The averaged values show the expected behavior: (i) at low power, $p<p_0$ only the pumped mode is populated; (ii) at $p\sim p_0$ the phonon starts oscillating so that there is a transfer of population to the second polariton mode while the pumped mode saturates. Here we notice that, depending on the size of the initial condition for the phonon, there might be an overshoot in the amplitude of $\langle|\phi_0(t)|^2\rangle$, which is reminiscent of the observed overshot in the experimental data shown in Fig.~\ref{Fig3}(c). This occurs for a small initial amplitude of the phonon and is to be expected as the systems is jumping between two different solutions with $A=0$ or $A\neq0$; (iii) for $p>2p_0$ all modes grow in intensity due to synchronization effects \cite{Ramos2024} not included in our previous analysis.

Figures \ref{Fig6}(c) and \ref{Fig6}(d) show the corresponding results for a system of seven polariton modes, $\phi_j(t)$ $j=-3,\dots,3$ with energies $\varepsilon_j=\varepsilon_0+j\,\delta\varepsilon$, coupled by a single mode of frequency $\Omega=3\delta\varepsilon/\hbar$. Only the zeroth level is pumped. Here we assume that such zeroth mode is more strongly coupled to the modes below ($J_{jk}=J_{01} (1+\delta_{0j})/2$ and $J_{kj}=J_{jk}$ with $k>j$). This is important to avoid the suppression of the phonon stimulation due to interference between modes at $\pm\hbar\Omega$. It is apparent from the figure that the general phenomenology is similar to the case of the two mode system but, quite importantly, only the modes separated by the phonon frequency from the pumped mode get a sizable population.
In fact, simulations (not shown) with $\Omega=2\delta\varepsilon/\hbar$ show that only the $j=\pm2$ modes are populated as expected.

Based on this, one expects that the addition of multiple phonon modes ($20$GHz, $60$GHz, $100$GHz) would lead to different `cascades' of the corresponding energy, each one signaling the mentioned features associated to the possibility of overshooting of certain intensities and the increase of the population of nearby (near resonant) modes as the corresponding phonon is stimulated. Yet, modeling such multi-mode polariton and phonon system, including the nonlinearities arising from interactions, while required to explain the full features observed in the experimental data, is a formidable task that we leave for future work.
%%%%%%%%%%%%%%%%%%%%%%%%%%%%%%%%%%%%%%%%%%%%%%%%%%%%%%%%%%%%%
% \paragraph*{\textbf{Discussion and outlook}} 
\section{Discussion and outlook}
%%%%%%%%%%%%%%%%%%%%%%%%%%%%%%%%

We have presented a practical implementation of the concept of a QCPL based on the sequential cascade of polariton bosonic condensates through a ladder of energy levels confined in a laterally microstructured wedged stripe. The shape of the structure is designed to display a parabolic-like effective potential with an almost energy-equidistant distribution of states localized along the stripe, with a large spatial overlap thus facilitating the optomechanical coupling. The level spacing is designed to fall close to the frequency of the fundamental vibrational mode $\nu_\mathrm{m}^{(0)} \sim 20$~GHz, with the polariton-reservoir and polariton-polariton non-linearities and dissipation all contributing to asynchronously lock them to optomechanical resonance when the phonon self-oscillation is induced. The asynchronous locking of both orbital and pseudospin states, the observation of spectral sidebands, and a strong harmonic time dynamics present in the time-delayed autocorrelation function $g^{(1)} (\tau)$, all contribute to firmly establish the presence of multimode phonon lasing involving the fundamental confined and vibrational overtones at $20$, $60$, and $100$~GHz. Analysis of the power dependence of the energy and intensity of the different modes allows us to determine both the condensation threshold $P_\mathrm{th}$, which is accompanied by phonon lasing involving the higher frequency phonons at $60$ and $100$~GHz, and at a slightly larger excitation power a second threshold with phonon lasing at the fundamental confined phonon frequency $\sim 20$~GHz. Interestingly, the stimulation of phonon emission leads to the distribution of spectral weight between the involved polariton states, which at the same time facilitates the relaxation of polaritons from the reservoir and towards the lower energy states of the structure. This can be an important effect for increasing the efficiency of vertically emitting polariton lasers and multi-wavelength polariton laser arrays.

From the birth of phonon lasing research, a variety of potential applications have been envisaged. These include the 3D imaging of structures in opaque environments with nanometer resolution, including single living cells~\cite{Danworaphong2015,Dehoux2015}, and also the ultrafast modulation of light which is of particular interest in microwave photonics applications involving super-high ($20$-$100$~GHz) frequencies~\cite{Lai2024}. While in principle the emission of phonons occurs mainly perpendicular to the DBRs and into the substrate, in-plane propagation and phonon manipulation useful on planar integrated applications has already been shown to exist in this platform~\cite{Xiang2024}.  The strong vibrational amplitudes accessible to phonon lasers also open the path to nonlinear phononics~\cite{Hoegen2018,Henstridge2022,Ginsberg2023}, which could be even accessible with continuous wave optical excitation as we do here. We note that while our proposal based on GaAs materials requires low-temperature operation due to the relatively small exciton binding energies ($T < 100$~ K), alternative polariton optomechanic approaches have been proposed involving molecular systems that could operate at room temperature~\cite{Shishkov2024}. The principle of bosonic cascade phonon lasing we demonstrate here could be also applied for the  spatially dependent ultra-fast temporal modulation of photonic lattices with applications in synthetic magnetic fields and non-reciprocal transport~\cite{Sounas2017,Galiffi2019}. In this context optomechanical applications have been theoretically proposed~\cite{Ruesnik}, including mechanically self-stimulated implementations as we show here with potential for dynamical gauge field simulators~\cite{Walter2016}. Quantum statistics and correlations of bosonic cascades are also intriguing developments, with potential applications in information and imaging technologies, that could be investigated with the material platform demonstrated here~\cite{Liew2016,Palomo2025}

%Polaromechanical lattices can be conceived as metamaterials that allows to tame polariton nonlinearities using well-defined phonon degrees of freedom~\cite{Polaromechanics2023}. 

%\paragraph*{Online content:}
%Any methods, additional references, Nature Research reporting
%summaries, source data, extended data, supplementary information, acknowledgements, peer review information; details of author
%contributions and competing interests; and statements of data and
%code availability are available at https://doi.org/.\\

%\section*{Data availability}
%The source data that support the findings of this study are available from the corresponding author upon reasonable request. All these data are directly shown in the corresponding figures without further processing.

%%%%%%%%%%%%%%%%%%%%%%%%%%%%
\begin{acknowledgments}
We acknowledge partial financial support from the ANPCyT-FONCyT (Argentina) under grants PICT 2019-0371, PICT 2020-3285, and SECTyP UNCuyo 06/C053-T1.
ASK and PVS acknowledge the funding from German DFG (grant 359162958). The authors thanks G. Rozas and S. Anguiano for valuable help in some of the optical experiments. Computational resources were provided by the HPC cluster of the Physics Department at Centro At\'omico Bariloche (CNEA), which is part of Argentina's National High-Performance Computing System (SNCAD).
\end{acknowledgments}
\appendix
%%%%%%%%%%%%%%%%%%%%%%%%%%%%%%%%%%%%%%%%%%%%%%%%%
\section{Polaritons on a wedged wire\label{ap:gGPE}}
%%%%%%%%%%%%%%%%%%%%%%%%%%%%%%%%%%%%%%%%%%%%%%%%%
It is well known that micro-cavity polariton condensates are well described by a generalized Gross-Pitaesvkii equation (gGPE) given by~\cite{Wouters2007} 
%%%%%%%%%%%%%%%%%%%
\begin{eqnarray}
\nonumber
 i \hbar \frac{\partial\psi(\bm{r},t)}{\partial t} &=& \left[- \frac{\hbar^{2}}{2 m} \bm{\nabla}^{2} + V(\bm{r})  \right] \psi(\bm{r},t) \\
 \nonumber
 &&+(g_R\,n(\bm{r})+g_p|\psi(\bm{r},t)|^{2})\psi(\bm{r},t)\\
 \label{gGPE1}
 && + \frac{i \hbar}{2}(R \,n(\bm{r},t)-\gamma)\psi(\bm{r},t)\,,\\
\nonumber
\frac{\partial n(\bm{r},t)}{\partial t} &=& P(\bm{r})-\left[\gamma_{R}+ R\,|\psi(\bm{r},t)|^{2}  \right] n(\bm{r},t) \,.\\
 \label{gGPE2}
 \end{eqnarray}
%%%%%%%%%%%%%%%%%%%
Equation \eqref{gGPE1} describes the dynamics of the lower polariton (LP) field $\psi(\bm{r},t)$ while Eq.~\eqref{gGPE2} describes the evolution of the exciton reservoir density $n(\bm{r},t)$. Here, $V(\bm{r})$ is the effective confinement potential for the LPs (wedged wire), $m$ is the effective mass, $g_R$ and $g_p$ are the polariton-reservoir and polariton-polariton interaction coupling constants, respectively, $R$ is the stimulated gain rate and $\gamma$ and $\gamma_R$ the polariton and reservoir losses, respectively.
The stationary off-resonance laser pump is described by $P(\bm{r})$, which here is assumed to have a Gaussian profile.
%%%%%%%%%%%%%%%%%%%%%%%%%%%%%%%%%%%%%%%%%%%%%%%%%
\subsection{2D confinement}
%%%%%%%%%%%%%%%%%%%%%%%%%%%%%%%%%%%%%%%%%%%%%%%%%
To better grasp the relevant features of the stationary solutions for this geometry, we initially consider only Eq. \eqref{gGPE1} neglecting gain, losses, and polariton-polariton interactions. Under these conditions, the solutions reduce to the energy eigenstates \(\psi_n(\bm{r}, t) = \psi_n(\bm{r}) e^{-\ci E_n t / \hbar}\) of the stationary Schr\"odinger equation
%%%%%%%%%%%%%%%%%%%%%%%%%%%%%%
\be
 \left[- \frac{\hbar^{2}}{2 m} \bm{\nabla}^{2} + V(\bm{r})  \right] \psi_n(\bm{r}) = E_n \psi_n(\bm{r})\,.
 \label{ap:eq:schroedinger}
\ee
%%%%%%%%%%%%%%%%%%%%%%%%%%%%%%
The effective mass is derived from the effective masses of the exciton, $m_X$, and of the cavity photon $m_C$, by using the relation  $m^{-1} = (1 - |X|^2){m_C}^{-1} + |X|^2 {m_X}^{-1}$ where $|X|^2$ is the so-called Hopfield coefficient, which depends on the detuning between the exciton and cavity photon modes. For our sample, $m_C = 4.2 \times 10^{-5} m_0$ and $m_X = 0.5\, m_0$, with $m_0$ the free-electron mass, and $|X|^2=0.2$. We use the following confinement potential
%%%%%%%%%%%%%%%%%%%%%%%%%%%%%%%%%
\be
  {V(x,y)}= V_0 \left[1-\mathcal{V}(x,x_\text{base}-x_\text{tip}) \mathcal{V}(y,w_y(x)) \right]\,,   
\ee
%%%%%%%%%%%%%%%%%%%%%%%%%%%%%%%%%
where we introduced the functions
%%%%%%%%%%%%%%%%%%%%%%%%%%%%%%%%%
\be
 \mathcal{V}(\alpha,w)= \frac{1}{2} \left[ \operatorname{erfc} \left( \frac{\alpha - \frac{1}{2}w}{\sqrt{2}\delta} \right) - \operatorname{erfc} \left( \frac{\alpha + \frac{1}{2}w}{\sqrt{2}\delta} \right) \right]\,,
\ee
%%%%%%%%%%%%%%%%%%%%%%%%%%%%%%%%%
and
%%%%%%%%%%%%%%%%%%%%%%%%%%%%%%%%%
\be
 w_y(x)=\frac{ (x_\text{base}-x) w_{y,\text{tip}} +  (x-x_\text{tip}) w_{y,\text{base}} }{x_\text{base}-x_\text{tip}}\,,
\ee
%%%%%%%%%%%%%%%%%%%%%%%%%%%%%%%%%
to properly account for the 2D shape of the stripe trap. Here $V_0=6$ meV is the estimated potential value outside the trap (etched region), taken with respect to the $1525$~meV of the LP planar cavity mode in the non-etched region. We take $\delta=0.15\,\mu$m, the transition length from the trap to the barrier in the potential profile $\mathcal{V}(\alpha,w)$. This value is consistent with the one used previously in Ref.~\cite{Kuznetsov2018} for similar samples. A qualitative account of the effect of the laser spot can be obtained by adding a Gaussian potential $V_\text{spot}(x,y)$---see below for the case of the 1D effective model. To model the $L=80~\mu$m long stripe we set $x_\text{base}=-x_\text{tip}=40 \mu$m as the two edges of the trap in the longitudinal direction. The width $w_y(x)$ changes linearly with $x$ between the tip and base positions with $w_{y,\text{tip}}=0.5\,\mu$m and $w_{y,\text{base}}=2\,\mu$m.  The results presented in Fig.~\ref{Fig2}(a) were obtained by finding the eigenstates of the system following a customary discretization of the 2D space and applying the finite difference method to solve Eq.~\eqref{ap:eq:schroedinger}.  
%%%%%%%%%%%%%%%%%%%%%%%%%%%%%%%%%%%%%%%%%%%%%%%%%
\subsection{Effective 1D problem}
%%%%%%%%%%%%%%%%%%%%%%%%%%%%%%%%%%%%%%%%%%%%%%%%%
Due to the fact that the variation of the width of the trap is smooth, $dw_y(x)/dx\ll1$, the system can be well described by an effective 1D problem. In order to do that, and since we are mainly interested on the lower energy part of the spectrum, we can safely consider a hard wall potential in the $x$ direction. On the contrary, in the transverse direction it is important to consider the finite height of the trap potential to avoid overestimating the effect of the confinement. With these assumptions, we introduce the following ansatz for the eigenfunction,  $\psi_{n,m}(\bm{r})=\phi_{n,m}(x)\chi_m(y,\{x\})$ where $\chi_m(y,\{x\})$ is the `local' transverse wave function satisfying
%%%%%%%%%%%%%%%%%%%%%%%%%%%%%%%%%
\begin{equation}
\left[-\frac{\hbar^2}{2m}\frac{\partial^2}{\partial y^2}+\tilde{\mathcal{V}}(y,\{x\})\right]\chi_m(y,\{x\})=\varepsilon_m(x)\chi_m(y,\{x\})\,,
\label{ap:eq:tranverse}
\end{equation}
%%%%%%%%%%%%%%%%%%%%%%%%%%%%%%%%%
where $\tilde{\mathcal{V}}(y,\{x\})=V_0(1-\mathcal{V}(y,w_y(x))$ is the effective transverse confinement potential at position $x$ and $\varepsilon_m(x)$ is the energy of the `local' transverse mode $m$. The corresponding equation for $\phi_{n,m}(x)$, neglecting all terms proportional to $dw_y(x)/dx$, is given by
%%%%%%%%%%%%%%%%%%%%%%%%%%%%%%%%%
\begin{equation}
\left[-\frac{\hbar^2}{2m}\frac{\partial^2}{\partial x^2}+\varepsilon_m(x)\right]\phi_{n,m}(x)=E_{n,m}\phi_{n,m}(x)\,,
\label{ap:eq:effective1D}
\end{equation}
%%%%%%%%%%%%%%%%%%%%%%%%%%%%%%%%%
where $\varepsilon_m(x)$ enters as an effective potential. To obtain an analytical expression for $\varepsilon_m(x)$ we first approximate $\tilde{\mathcal{V}}(y,\{x\})$ by a square well potential. In that case, one can obtain an analytic expression for the eigenvalues. For instance, the even modes (with respect to $y$) are given by the relation $\sqrt{\tilde{V}_0}=\sqrt{\tilde{\varepsilon}_m}/|\cos(\sqrt{\tilde{\varepsilon}_m}\pi/2)|$ with $\tilde{V}_0=V_0 8L^2m/h^2$ and $\tilde{\varepsilon}_m=\varepsilon_m8L^2m/h^2$ that needs to be solved for $\tilde{\varepsilon}_m$. For simplicity, we limit ourselves here to consider only the first transverse mode, for which an approximate expression for the energy can be obtained,
%%%%%%%%%%%%%%%%%%%%%%%%%%%%%%%%%
\begin{equation}
\varepsilon_0(x)\simeq\epsilon(x)\left(1+\frac{2}{\pi}\sqrt{\frac{\epsilon(x)}{V_0}}\right)^{-2}\,,
\label{ap:eq:effective_energy}
\end{equation}
%%%%%%%%%%%%%%%%%%%%%%%%%%%%%%%%%
with $\epsilon(x)=\hbar^2\pi^2/2mw_y^2(x)$. The spatially resolved spectrum obtained by solving Eq.~\eqref{ap:eq:effective1D} using Eq.~\eqref{ap:eq:effective_energy} is shown in Fig.~\ref{Fig2}(a).
%, where the solid white line corresponds to $\varepsilon_0(x)$. Note that close to the wider edge the effective potential is nearly linear leading to eigenvalues and eigenfunctions similar to those of a triangular potential. In addition, 
The spacing between energy levels in our sample is roughly constant, up to approximately the $50$th level, and of the order of the energy $\nu_m^{(0)}$ of the $20$ GHz cavity phonon (see Fig.~\ref{Fig2}(c))---the finite value of $V_0$ is important for this as the hard wall confinement leads to a considerable larger level spacing. 
%%%%%%%%%%%%%%%%%%%%%%%%%%%%%%%%%%%%%%%%%%%%%
\begin{figure}[t]
    \centering
    \includegraphics[width=0.95\linewidth]{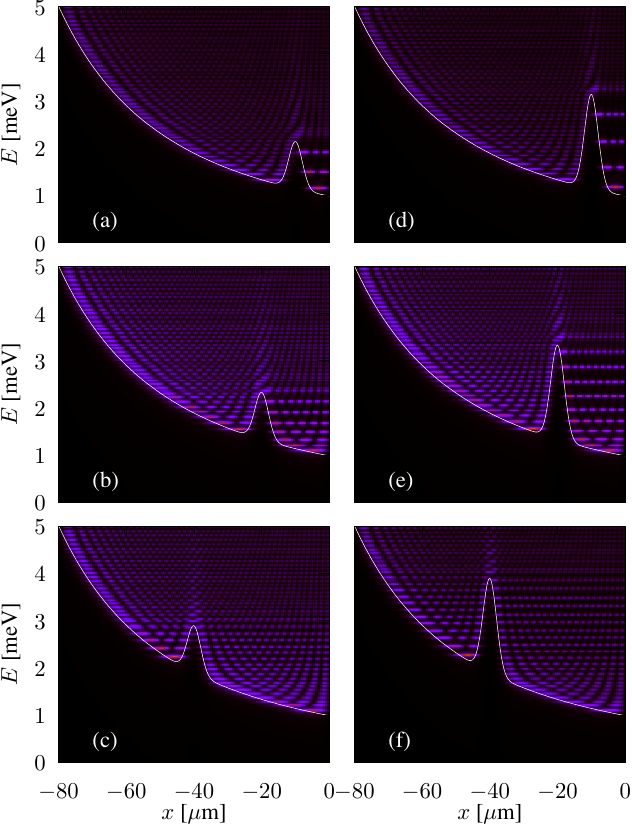}
    \caption{Spatially resolved spectral density obtained by solving Eq.~\eqref{ap:eq:effective1D} with $V_\text{tot}(x)$ in place of $\varepsilon_0(x)$. Here $\sigma=2\mu$m, $P_0=1$ meV (a-c) and $P_0=2$ meV (d-c) while $x_0=10\,\mu$m, $20\,\mu$m and $40\,\mu$m from top to bottom. The position and power dependent confinement induced by the laser pump is apparent from the figure.}
    \label{ap:fig:1}
\end{figure}
%%%%%%%%%%%%%%%%%%%%%%%%%%%%%%%%%%%%%%%%%%%%

It is straightforward to add, within the same approximation, the effect of the laser pump (at low power) by simply adding to the effective confinement potential $\varepsilon_0(x)$ a local (Gaussian-like) potential whose amplitude is proportional to the laser power, $V_\text{spot}(x)=g_R n_I(x)=P_0 \exp[-\frac{(x-x_0)^2}{2\sigma^2}]$, where $n_I(x)$ is the inactive reservoir. The effect of such potential is illustrated in Figs.~\ref{ap:fig:1}(a-c) for $P_0=1$~meV, $\sigma=2\,\mu$m and $x_0=10\,\mu$m, $20\,\mu$m and $40\,\mu$m, respectively. The total potential profile, $V_\text{tot}(x)=\varepsilon_0(x)+V_\text{spot}(x)$, in indicated by the (white) solid line as a reference. We observe that the presence of the pump  leads to a two-trap structure with a quite different energy level separation depending on the position $x_0$ of the laser spot. Clearly, depending on its position, the laser can  lead to a strong confinement on the wider size of the trap. This in turns allows to tune the energy levels separation of the states located on that side of the trap. The level spacing on the left side is much less affected due to the spatial dependence of $\varepsilon_0(x)$. This effect causes an intercalation of the energy levels located on each side of the pump. Figures \ref{ap:fig:1}(c-f) illustrate the effect for the same positions but a higher value of $P=2$~meV.
%%%%%%%%%%%%%%%%%%%%%%%%%%%%%%%%%%%%%%%%%%%%%
\begin{figure}[t]
    \centering
    \includegraphics[width=1\linewidth]{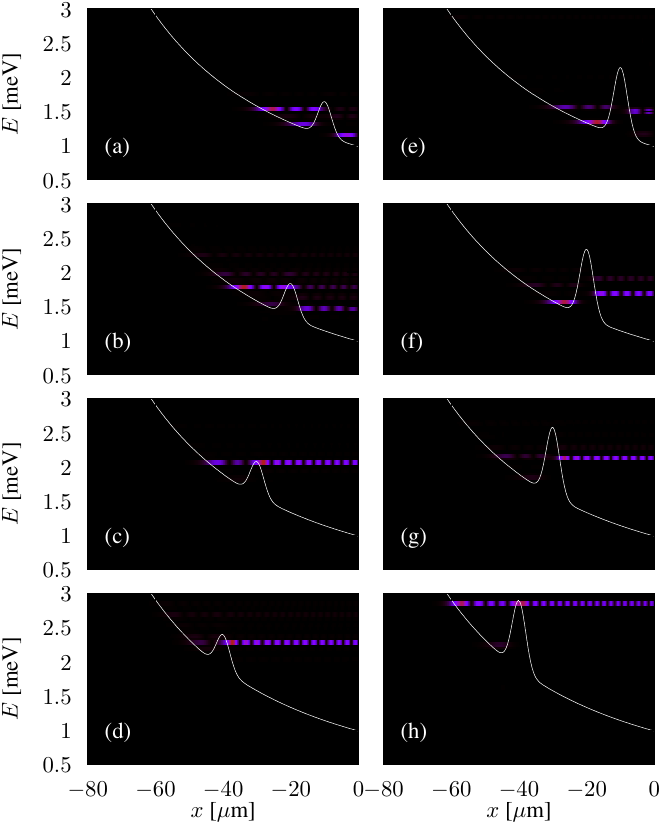}    \caption{Spatially resolved spectral density obtained by solving Eq.~\eqref{ap:eq:gGPE_effective}. Here $\sigma=2\mu$m, $P_0=0.5$ meV (a-d) and $P_0=1$ meV (e-f) while $x_0=10\,\mu$m, $20\,\mu$m, $30\,\mu$m and $40\,\mu$m from top to bottom. We used $\sigma_1=2\sigma$ and $P_1\gg1$. The solid (white) line indicates the total effective potential $ V_\mathrm{tot}(x)$,The color scale is linear and independently set on each panel. Note that in most cases only few modes are dominant or even a single one.}
    \label{ap:fig:2}
\end{figure}
%%%%%%%%%%%%%%%%%%%%%%%%%%%%%%%%%%%%%%%%%%%%

Based on these results we now construct an effective $1$D version of Eq.~\eqref{gGPE1} as follows
%%%%%%%%%%%%%%%%%%%%%%%%%%%%%%%%%%%%%%%%%%%%%
\begin{eqnarray}
\nonumber
 i \hbar \frac{\partial\psi(x,t)}{\partial t} &=& \left[- \frac{\hbar^{2}}{2 m} \frac{\partial^2}{\partial x^2} + V_\mathrm{tot}(x) \right] \psi(x,t) \\
 && + \frac{i \hbar\gamma}{2}(\tilde{n}_A(x,t)-1)\psi(x,t)\,.
  \label{ap:eq:gGPE_effective}
\end{eqnarray}
%%%%%%%%%%%%%%%%%%%%%%%%%%%%%%%%%%%%%%%%%%%%%
Here $\tilde{n}_A(x,t)= p(x)/(1+|\psi(x,t)|^2)$ represents the (normalized) gain due to the  active reservoir, including the saturation effect, obtained from Eq.~\eqref{gGPE2} and requiring $\partial n_A(x,t)/\partial t\sim0$ (instantaneous response of the reservoir). We take $p(x)=P_1\exp[-\frac{(x-x_0)^2}{2\sigma_1^2}]$ with $\sigma_1>\sigma$ to account for the diffusion of excitons in the active reservoir. 

Figure \ref{ap:fig:2} shows some examples of the stationary solutions obtained by the numerical integration  of Eq.~\eqref{ap:eq:gGPE_effective} using standard ODE solvers. We find that, typically, only a few modes, or even a single mode, are populated. Which mode/s is/are selected is strongly dependent on the position ($x_0$) of the laser pump.  This sensitivity with respect to $x_0$ is also found in the  experimental data. 

Clearly, in order to describe the emergence of the phonon cascade, as observed in the experiments, an optomechanical coupling between polariton and cavity phonons must be added as discussed in the main text using a simple model.
\section{Experimental evidence of the effect of the laser pump on the potential and modes of the wedged stripe}
\begin{figure*}[t]
    \centering
    \includegraphics[width=1\linewidth]{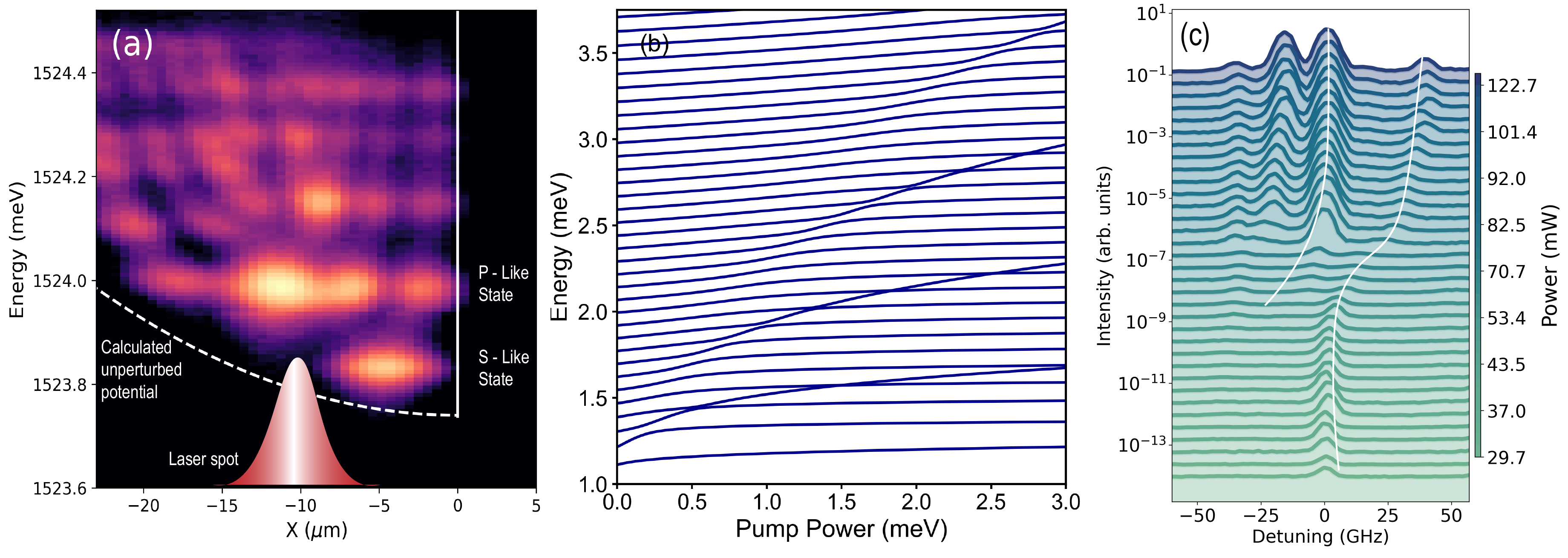}
    \caption{
    \textbf{Polariton states in a wedged stripe with a laser pump.}
\textbf{(a)} Detail of spectrally resolved spatial image for the region close to the laser spot localized at $\sim 8 \mu$m from the wide edge of the stripe (indicated with white lines). The calculated wedge stripe potential is indicated with a dashed curve. \textbf{(b)} Calculated variation of the energy of the confined polariton modes with laser power. Modes to the left and to the right of the laser spot depend differently with power, and are coupled through the pump-induced barrier, leading to avoided crossings. An experimental example of such avoided crossing is illustrated in \textbf{(c)}. 
    }
    \label{ap: ExPumpEffect}
\end{figure*}

%%%%%%%%%%%%%%%%%%%%%%%%%%%%%%%%%%%%%%%%%%%%%%%
We discuss here in more detail the experimental consequences of the perturbation induced by the laser spot on the effective potential sensed by the trapped polaritons (calculation shown in Appendix A). The laser excitation leads to a local blue shift of the polariton modes. This blue-shift represents a barrier that effectively defines two coupled trapping potentials along the stripe: one to the left of the laser spot, limited by the parabolic-like potential due to confinement, and the other to its right, limited by the stripe end. The effective confinement potential limited to the right by the stripe's end is particularly notable for spots close to the stripe wider limit, as shown in Fig.~\ref{ap: ExPumpEffect}(a). This experiment corresponds to the wedged $80 \mu$m long stripe, with lateral dimension varying linearly from $0.5 \mu$m to $2 \mu$m, and with the laser spot located at $\sim 8 \mu$m from the wider edge of the stripe. In this case one can identify, between the pump and the stripe edge, confined states that display s- and p-like symmetry,  with the other higher energy modes corresponding to extended states. 

The existence of these two kinds of trapped modes, to the left and to the right of the laser spot, connected through the laser-induced barrier, leads to some peculiar phenomena. Figure \ref{ap: ExPumpEffect}(b) shows the calculated variation of some modes with laser power. Modes to the left and to the right of the laser-induced barrier depend differently on the pump power and, since they are coupled through the barrier, this leads to avoided crossings. An experimental example of such avoided crossings is presented in Fig.~\ref{ap: ExPumpEffect}(c), providing further evidence of the overall good understanding of the character of the modes of the wedged stripe.

%%%%%%%%%%%%%%%%%%%%%%%%%%%%%%%%%%%%%%%%%%%%%
%%%%%%%%%%%%%%%%%%%%%%%%%%%%%%%%%%%%%%%%%%%%

%\bibliography{refe}

%Corresponding author: A.F., alejandro.fainstein@ib.edu.ar

%\begin{references}

%\section*{Acknowledgements} 

%\section*{Author contributions}
%D.L.C. and A.S.K. have contributed equally.  D.L.C, I.P. and A.E.B. contributed to the pump and probe and locking experiments, and A.S.K to the sample characterization and measurement of the polariton dispersions.  A.S.K., K.B., and P.V.S. designed and fabricated the structured microcavity sample.  P.S., A.A.R., G.U, A.E.B., and A.F. outlined theoretical aspects,  with A.A.R and G.U developing the synchronization theory and performing the related numerical simulations.  All authors contributed to the discussion and analysis of the results.  P.V.S. and A.F. conceived and directed the project. A.F. prepared the manuscript with inputs from all co-authors.

%\section*{Competing interests}
%The authors declare no competing interests.

%\section*{Additional information}

%\paragraph*{\textbf{Supplementary information}} is available for this paper at https://doi.org/... \\

%\paragraph*{\textbf{Correspondence}} and requests for materials should be addressed to A.F.
\end{document}